\documentclass[usenatbib]{mn2e}
\usepackage{graphicx}
\usepackage{color}

\title[SMSS J130522.47-293113.0: a high-latitude stellar X-ray source with pc-scale outflow relics?]
{SMSS J130522.47-293113.0: a high-latitude stellar X-ray source with pc-scale outflow relics?}
\author[Da Costa et al.]
{G. S. Da Costa,$^1$ R. Soria,$^{2,3,4}$ S. A. Farrell,$^{4,5}$  D. Bayliss,$^{1,5}$ M. S. Bessell,$^1$ 
\newauthor
F. P. A. Vogt,$^{1,6}$ G. Zhou,$^{1,7}$ S. D. Points,$^8$ T. C. Beers,$^9$ \'{A}. R. L\'{o}pez-S\'{a}nchez,$^{10}$    
\newauthor
K. W. Bannister,$^{11}$ M. Bell,$^{4,12}$ P. J. Hancock,$^{3,4}$ D. Burlon,$^4$ B. M. Gaensler,$^{4,5,13}$    
\newauthor
E. M. Sadler,$^{4,5}$ S. Tingay,$^3$ S. C. Keller,$^1$ B. P. Schmidt,$^{1,5}$ and P. Tisserand$^1$\\
\\
$^1$Research School of Astronomy and Astrophysics, Australian National 
University, Canberra, ACT 0200, Australia\\
$^2$National Astronomical Observatories, Chinese Academy of Sciences, Beijing 100012, China\\
$^3$International Centre for Radio Astronomy Research, Curtin University, Perth, WA 6845, Australia\\
$^4$Sydney Institute for Astronomy, School of Physics, The University of Sydney, Sydney, NSW 2006, Australia\\
$^5$ARC Centre of Excellence for All-sky Astrophysics (CAASTRO)\\
$^{6}$Observatoire de Gen\`{e}ve, Universit\'{e} de Gen\`{e}ve, Chemin des maillettes 51, 1290, Sauverny, Switzerland\\
$^7$European Southern Observatory, Av.\ Alonso de C\'{o}rdova 3107, 763 0355 Vitacura, Santiago, Chile\\
$^8$Harvard-Smithsonian Center for Astrophysics, 60 Garden St, Cambridge, MA 02138, USA\\
$^9$Cerro-Tololo Inter-American Observatory, Casilla 603, La Serena, Chile\\
$^{10}$Department of Physics and JINA Center for the Evolution of the Elements, University of Notre Dame, Notre Dame, IN 46556, USA\\
$^{11}$Australian Astronomical Observatory, PO Box 915, North Ryde, NSW 1670, Australia\\
$^{12}$CSIRO Astronomy \& Space Sciences, PO Box 76, Epping, NSW1710, Australia\\
$^{13}$University of Technology Sydney, 15 Broadway, Ultimo NSW 2007, Australia\\
$^{14}$Dunlap Institute for Astronomy and Astrophysics, The University of Toronto, ON M5S 3H4, Canada}
 
\begin{document}

\maketitle

\begin{abstract}

We report the discovery of an unusual stellar system SMSS J130522.47-293113.0.  The optical spectrum 
is dominated by a blue continuum together with emission lines of hydrogen, neutral and ionized helium, 
and the N {\sc iii}, C {\sc iii} blend at $\sim$$\lambda$4640--4650\AA.   
The emission line profiles vary in strength and position on timescales as short as 1 day, while 
optical photometry reveals fluctuations of as much as $\sim$0.2 mag in $g$ on timescales as short 
as 10--15 min.  The system is a weak X-ray source ($f_{0.3-10} = (1.2 \pm 0.1) \times 10^{-13}$ ergs 
cm$^{-2}$ s$^{-1}$ in the 0.3--10 keV band) but is not detected at radio wavelengths (3$\sigma$ upper 
limit of 50$\mu$Jy at 5.5GHz).  The most intriguing property of the system, however, is the existence of 
two ``blobs'', a few arcsec in size, that are symmetrically located 3$^\prime$.8 (2.2 pc for our preferred 
system distance of $\sim$2 kpc) each side of the central object.  
The blobs are detected in optical and near-IR broadband images but do not show any 
excess emission in H$\alpha$ images.  We discuss the interpretation of the system, suggesting that the 
central object is most likely a nova-like CV, and that the blobs are relics of a pc-scale accretion-powered 
collimated outflow.

\end{abstract}

\begin{keywords}
stars: activity; stars: emission-line; stars: jets; X-rays: binaries
\end{keywords}

\section{Introduction} \label{Intro} 

Serendipity plays a significant role in astrophysics through the discovery of unexpected objects in surveys, 
and such discoveries often result in the opening up of entirely new areas of inquiry.  Within the field of 
compact mass-accreting binary stellar systems an example is the unanticipated nature of the
object SS433 in the compilation of \cite{SS77}, a catalogue of 
H$\alpha$ emission line objects derived from objective prism surveys of the Milky Way.
Follow-up observations of the source \citep[e.g.,][]{CM78} revealed a complex optical spectrum, 
which is now 
understood as arising in an eclipsing high mass X-ray binary system with two oppositely directed,
precessing, and continuously emitting relativistic jets of gas \citep[][and references therein]{KB11}.  Because of the 
similarity of these properties with those of quasars, albeit at a much lower mass scales, SS433 has become the 
prototype of the class of Galactic objects known as microquasars: X-ray binaries with associated mass outflow 
in the form of bipolar jet structures that are generally seen at radio wavelengths.  

The jet structures seen in microquasars are only one example of jet phenomena associated with stellar objects.  For example, accretion driven
collimated jets are also seen in young stellar objects, in both low and high mass X-ray binaries, and in a number of systems involving accreting white dwarfs \citep[e.g.,][and references therein]{KR08}.
What is much rarer, however, is the existence of the relics of jets in stellar systems, that is, a situation where the source of the jet power has `turned-off' but where direct evidence supporting the previous existence of jets or outflows remains. 

Here we report the  discovery of two unusual patches of diffuse emission symmetrically located either side of a bright optical-line-emitting source.  The system was serendipitously observed as part of the AEGIS spectroscopy survey (P.I.\ Keller) of candidate Galactic halo stars.  We argue that the central source is an accreting white dwarf, specifically a nova-like cataclysmic variable (CV), at a distance of 
$\sim$2 kpc.  The two diffuse ``blobs'' are interpreted as 
relics of now defunct pc-scale collimated outflows from the central accreting source.  If this is the case it would confirm that accreting white dwarfs are capable of producing large-scale collimated outflows or jets, similar to those seen in neutron star and black hole accretion stellar systems  \citep[e.g.,][and references therein]{KR08}.   Nevertheless, we cannot rule out the possibility that the ``blobs'' are 
remnants related to a historical nova outburst from the central source.

The discovery observations are described in the following section.  In section 3 a series of follow-up observations at wavelengths from radio to X-ray are described, while the possible interpretations of the central object are explored in section 4.  The potential relic jets are discussed in section 5 and the results are summarised in section 6.  

\section{Discovery Observations}


The $AEGIS$ survey (P.I.\ Keller) 
was a moderate scale (4 semesters, 15 nights per 
semester) spectroscopic survey carried out with the AAOmega multi-fibre spectrograph system 
\citep[e.g.,][]{RS06} at the Anglo-Australian Telescope.  The aim of the survey is to
constrain the chemodynamical evolution of Milky Way through the study of the kinematics and abundances of 
stars in the halo and the outer thick disk.  The input catalogue for the spectroscopic observations was derived from 
photometry of approximately 200 two-degree diameter fields obtained during the commissioning of the SkyMapper 
telescope \citep{K07} at Siding Spring Observatory.  In particular, the gravity and metallicity sensitivity of the 
SkyMapper photometric system \citep{K07} allows the focus of the target list to be on blue horizontal branch (BHB), 
red clump, and metal-poor star candidates.  The observations were made with the 580V grating in the blue arm of 
the spectrograph, providing a wavelength coverage of approximately  $\lambda$3750--5400\AA\/ 
with a resolving power {\it R} = $\lambda/\Delta\lambda$ of $\sim$1300.  The red arm used the 1700D grating centred at $\lambda$8600\AA\/  allowing coverage of the
Ca {\sc ii} triplet region with {\it R} $\approx$ 10,000.   All of the survey observations are run through a uniform 
data reduction process, but the availability of the reduction code 
{\sc 2dfdr}\footnote{www.aao.gov.au/AAO/2df/aaomega/aaomega\_software.html\\\#drcontrol}
at the telescope allows a quick analysis, permitting a visual inspection of the spectra and thus the immediate
identification of any unusual objects.  

\begin{figure}
\centering
\includegraphics[angle=-90.,width=0.47\textwidth]{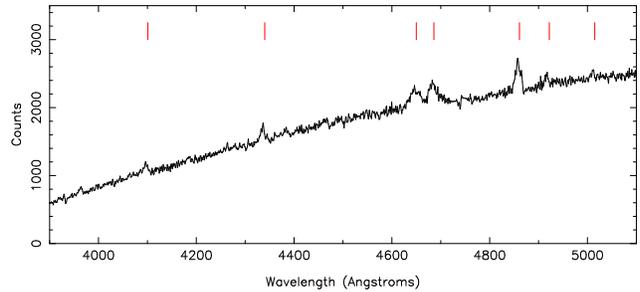}
\caption{Discovery spectrum of the unusual star {\it SMSS J1305--2931} from the AAT AAOmega blue camera 
observation.  The spectrum has not been flux or continuum corrected.  The Balmer lines H$\beta$, $H\gamma$ 
and H$\delta$ are clearly visible in emission as are the He~{\sc i} lines at $\lambda$4922\AA\/ and 
$\lambda$5015\AA.  Also present is N~{\sc iii} and C~{\sc iii} emission at $\lambda$4640--4650\AA\/ and He {\sc ii} emission 
at $\lambda$4686\AA. 
\label{Sp_Fig1}}
\end{figure}

Figure \ref{Sp_Fig1} shows the reduced AAOmega blue arm spectrum of one such unusual object, which was included in the survey input catalogue as a candidate BHB star on the basis of its blue colour.  We denote the star
by its SkyMapper survey DR1.1\footnote{Wolf et al., (2018); DOI: 10.4225/41/593620ad5b574}
designation of SMSS J130522.47--293113.0 (hereafter SMSS J1305--2931).  The AEGIS field containing 
SMSS J1305--2931 was 
observed on 2012 April 13 (UT), and the spectrum shown in Fig.\ \ref{Sp_Fig1} is the combination of three 1500 s integrations taken in 
moderately good seeing (FWHM $\approx$ 1$^{\prime\prime}$.7; cf.\ on-sky fibre diameter 2$^{\prime\prime}$.0) and with clear skies.  The spectrum shows  broad emission in the Balmer lines, the 
characteristic blend of emission from N~{\sc iii} and C~{\sc iii} at $\lambda$ $\approx$ 4640--4650\AA, 
and He {\sc ii} at $\lambda$ $\approx$ 4686\AA\/ on a blue continuum.  There are no significant absorption lines present.  The 
peak of the H$\beta$ profile in Fig.\ \ref{Sp_Fig1} yields a heliocentric velocity of approximately --250 km s$^{-1}$.
The intrinsic FWHM of the line is of order 500 km s$^{-1}$ and multiple components may be present.  
The spectrum is suggestive of a previously uncatalogued compact accreting object in a binary system.  We note for completeness that the corresponding red spectrum is of low S/N and shows no readily identifiable features. 

\begin{figure}
\centering
\includegraphics[angle=0.,width=0.47\textwidth]{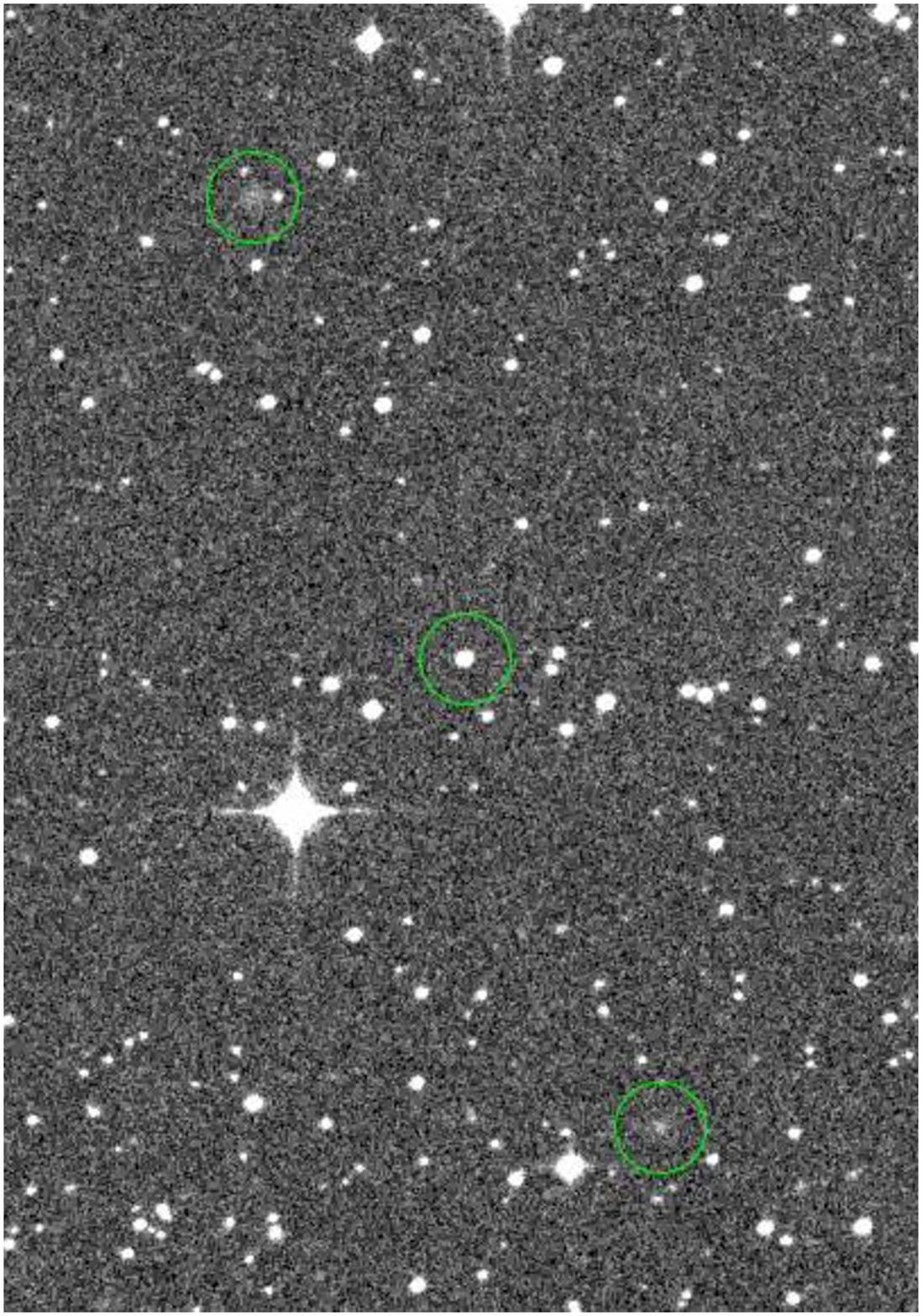}
\caption{DSS-Red image of the field of the unusual star SMSS J1305--2931.  The star is circled at the centre of the field; shown also with circles are two ``blobs'' that lie 
symmetrically $\sim$3$^{\prime}$.8 either side of the central star.  These faint features are also 
visible on the DSS-B image.  In the figure North is up and East is to the left, and the field shown is 
10$^{\prime}$
 in size in the N-S direction.
\label{DSS_Fig2}}
\end{figure}

The most intriguing aspect of this source, however, is the presence of two low surface brightness 
``blobs'' symmetrically located each side of the central star at angular distances of $\sim$3$^{\prime}$.8.
They were first noted on the Digitized Sky Survey (DSS) red image (Fig.\ \ref{DSS_Fig2}) but are also visible on the DSS blue image.  They are just visible in the Pan-STARRS1 \citep{FL16} $y/i/g$ image stack of the field, and are most evident on the $i$ image stack.  The blobs are $\sim$10$^{\prime\prime}$ in diameter,
and are of relatively uniform surface brightness without obvious internal structure, 
although the SW-component is perhaps more compact.

Their symmetrical location, together with the emission-line spectrum
of the central star, marks out SMSS J1305--2931 as a system of special interest.  Consequently, we have carried out a number of follow-up studies to investigate nature of the central source and the accompanying blobs, as possible interpretations include a previously unknown microquasar or some other type of accreting system with extended collimated outflows.

\section{Follow-up Studies}

\subsection{Radio Observations} \label{radio}

There is no radio source listed in the NVSS catalogue \citep{Condon98} near the position of 
SMSS J1305--2931, nor is there any detection at this position in the SUMSS catalogue \citep{Mauch03}.
We therefore carried out radio observations of the field of SMSS J1305--2931 at 5.5 and 9 GHz with the Australia Telescope Compact Array, as a `Target of 
Opportunity' (proposal ID CX242), during the period 2012 July 5--8.  No emission 
consistent with the location of the central star or of the blobs was detected at either frequency.  
The 3$\sigma$ upper limits are 50$\mu$Jy at 5.5GHz 
and 70$\mu$Jy at 9GHz.  There are a number of other point sources present in the field, but none are 
coincident with the optical positions of any of the components of the SMSS J1305--2931 system.

\subsection{Optical/IR imaging}

\subsubsection{Archival data}

We note that the central star lies at J2000 coordinates RA = $13^h05^m22^s.47$, Dec =  
$-29^{\circ}31^{\prime}13^{\prime\prime}.2$ in the Gaia DR1 catalog \citep{Gaia16} coincident with the SkyMapper DR1.1 position.  The corresponding Galactic coordinates are $l = 306^{\circ}.6$ and $b = 33^{\circ}.25$.  The reddening in this direction is relatively small: E($B-V$) = 0.11 mag \citep{SFD98} 
or 0.09 mag with the \citet{SF11} recalibration.   The central star is also listed in the UCAC4 \citep{Z12}, 
PPMXL \citep{R10}, SPM \citep{G11} and APOP \citep{Qi15} astrometric catalogues, and in each case the measured
proper motion is consistent with zero to within the errors, which are 3--6 mas yr$^{-1}$ in size.
Catalogue photometry for the central star, for wavelengths from the uv ({\it GALEX}) through to the IR (e.g., {\it WISE}), are given in Table \ref{Tab1}.  We note also that Table \ref{Tab1} lists the improved $JHK$ photometry discussed in \S\ref{NIRphot}, rather than the less precise values in the 2MASS point source catalogue.
The star is also in the Pan-STARRS1 data archive \citep{FL16}, with 
$g$ = 16.32, $g-r$ = --0.07 and $g-i$ = --0.24 mag on the Pan-STARRS1 system. 

\begin{table}
\begin{center}
\caption{Properties of the Central Star of the SMSS J1305--2931 system}
\label{Tab1}
\begin{tabular}{lll}
\hline
Parameter & Value & Source\\
\hline
\\
RA (2000) & 13:05:22.47 & 1\\
Dec (2000) & --29:31:13.0 & 1\\
Reddening E($B-V$) & 0.09--0.11 & 2,3\\
$V$ (APASS) & 16.52 $\pm$ 0.04 & 4\\
$B-V$ (APASS) & 0.09 $\pm$ 0.04 & 4\\
$g$ (SkyMapper)$^a$ & 16.32 $\pm$ 0.06 & 5\\
$u-g$ (SkyMapper) & 0.02 $\pm$ 0.12 & 5\\
$v-g$ (SkyMapper) & 0.15  $\pm$ 0.13 & 5\\
$g-r$ (SkyMapper) & --0.16 $\pm$ 0.09 & 5\\
$g-i$ (SkyMapper) & --0.30 $\pm$ 0.07 & 5\\
$K_{s}$ (AAT) & 15.84 $\pm$ 0.02 & 6\\
$J-K_{s}$ (AAT) & 0.23 $\pm$ 0.03 & 6\\
$H-K_{s}$ (AAT) & 0.13 $\pm$ 0.03 & 6\\
$W1$ (WISE 3.36 $\mu$m) & 15.79 $\pm$ 0.05 & 7\\
$W2$ (WISE 4.6 $\mu$m) & 15.89 $\pm$ 0.13 & 7\\
$FUV$ (GALEX $\lambda_{\rm eff}$ 1500 \AA) & 17.05 $\pm$ 0.04 & 8\\
$NUV$ (GALEX $\lambda_{\rm eff}$ 2300 \AA) & 16.70 $\pm$ 0.02 & 8\\
$uvw2$ ({\it Swift}/UVOT)$^a$ & $15.18 \pm 0.03 $ & 6\\
$uvm2$ ({\it Swift}/UVOT)$^a$ & $15.09 \pm 0.03 $ & 6\\
$uvw1$ ({\it Swift}/UVOT)$^a$ & $15.12 \pm 0.04 $ & 6\\
$u$ ({\it Swift}/UVOT)$^a$ & $15.40 \pm 0.03 $ & 6\\
$b$ ({\it Swift}/UVOT)$^a$ & $16.56 \pm 0.03 $ & 6\\
$v$ ({\it Swift}/UVOT)$^a$ & $16.46 \pm 0.03 $ & 6\\

\hline
$^a$ variable at the $\sim$0.2 mag level. & &\\
\multicolumn{3}{l}{1: \cite{Gaia16}}\\
\multicolumn{3}{l}{2: \cite{SFD98}}\\
\multicolumn{3}{l}{3: \cite{SF11}}\\
\multicolumn{3}{l}{4: \cite{AP12}}\\
\multicolumn{3}{l}{5: \cite{Wolf+17}}\\
\multicolumn{3}{l}{6: this work}\\
\multicolumn{3}{l}{7: \cite{CuW13}}\\
\multicolumn{3}{l}{8: \cite{BH12}}\\
\end{tabular}
\end{center}
\end{table}

\subsubsection{Faulkes-South Observations}







We used observations with the 2-m Faulkes-South telescope Spectral Camera CCD imaging system to provide optical magnitudes and colours for the blobs, and to constrain the variability of the central star.  The CCD camera, with 2$\times$2 pixel binning, provides images of a 
field $\sim$10$^{\prime}\times$10$^{\prime}$ in size at a scale of 0$^{\prime\prime}$.30 per binned pixel.
Images were centered on the central star and the field size is large enough to comfortably include 
both of the blobs.

The first set of observations were carried out in late April and May 2012.  The specific frames analyzed are a set of 3$\times$600 s images taken with the 
H$\alpha$ filter 
(FWHM 300\AA) on 2012 April 30, which were accompanied by 3$\times$180~s images with the $r$ filter, a 600~s $V$ 
exposure taken on 2012 May 1, and a set of 3$\times$600 s images taken with an [O{\sc iii}] 5007 filter
(FWHM 30\AA) on 2012 May 11.  The seeing for the H$\alpha$ and $r$-band images was poor 
($\sim$3$^{\prime\prime}$.2) while that for the $V$ and [O{\sc iii}]
observations was moderately good ($\sim$1$^{\prime\prime}$.7).  
The central star is well-detected on all the frames while the blobs are faintly visible on the 
combined $r$ frame and on the $V$ frame, may be present on the 
combined H$\alpha$ frame, and are not readily apparent on the combined [O{\sc iii}] frame.
In order to improve the sensitivity to low surface brightness features such as the blobs, each of the 
four frames (combined H$\alpha$, combined $r$, $V$ and combined [O{\sc iii}]) was block-averaged 
with a 4$\times$4 pixel box, and all subsequent photometry was carried out on these block-averaged frames.

A second set of observations was obtained with the same CCD camera in 2013 March.  In these 
observations the system was repeatedly imaged over three nights with 60~s exposures and the $g$ filter.
The time between exposures was 96~s.  The system was followed for $\sim$66 min on 2013 March 16 
(UT Date), $\sim$200 min on March 18, and $\sim$143 min on March 20.  For the first two nights the 
seeing was moderate (1$^{\prime\prime}$.8--2$^{\prime\prime}$.4) with intermittent cloud, while for the 
third night conditions were mostly clear with reasonable seeing (1$^{\prime\prime}$.4).

First, to study the properties of the blobs, a single $g$-band image was made by combining thirty
60~s integrations from 2013 March 20, which is the best seeing data.   The blobs are clearly present 
on the combined block-averaged image and, to a precision of $\sim$1$^{\prime\prime}$, the positions 
of both the NE 
and SW blobs on this image, given in Table \ref{Tab2}, are identical to those on the DSS images 
(and to the positions on the Pan-STARRS1 $i$-band image stack).  
The angular distance of the NE blob from the central star is 228$^{\prime\prime}$.4, 
while that for the SW blob is 228$^{\prime\prime}$.7.  Given that the centres of the blobs are difficult to 
measure to better than $\sim$1$^{\prime\prime}$ precision, it is clear that they are equidistant from the 
central star at a level better than 1 percent.  Similarly, the position angles of the two blobs relative to the 
central star are 24$^{\circ}$.3 for the NE blob and 180$^{\circ}$+22$^{\circ}$.3 for the SW blob, once 
more indicating a high degree of symmetry. However, despite the obvious symmetries, it is apparent 
from this frame
that the NE blob is somewhat more extended and less centrally concentrated than the SW blob.
This is also evident on the DSS images (see Fig.\ \ref{DSS_Fig2}) and from the $H$-band image
discussed in \S \ref{NIRphot} (see Fig.\ \ref{H_phot_fig}).

The $g$ and $r$ magnitudes for stars on the frames in the APASS 
catalogue \citep{AP12}\footnote{http://www.aavso.org/apass} were used to define a transformation 
from instrumental aperture photometry to the APASS system, with zero point uncertainties of 
0.02--0.03 mag.
Using this transformation the mean surface brightnesses of the NE and SW blobs, 
within a 3$^{\prime\prime}$ radius aperture that encompasses most of the light were determined.  
The values are given in Table \ref{Tab2}.  
The $r$ magnitudes for the blobs were measured in the same way on the combined $r$-band image, 
and they yield relatively red $(g-r)$ colours for the blobs, which are also given
in Table \ref{Tab2}.  

\begin{table}
\caption{Properties of the `blobs' in the SMSS J1305--2931 system}
\label{Tab2}
\begin{center}
\begin{tabular}{ll}
\hline
Parameter & Value\\
\hline
\multicolumn{2}{c}{{\it NE blob}} \\[2pt]
RA (2000) & 13$^h$05$^m$29$^s$.8 ($\pm$1$^{\prime\prime})$\\
Dec (2000) & --29$^{\circ}$27$^{\prime}$46$^{\prime\prime}$ ($\pm$1$^{\prime\prime})$\\
$g$ (APASS system) & 24.7 $\pm$ 0.1 mag arcsec$^{-2}$\\
$g-r$ (APASS system) & 1.15 $\pm$ 0.05\\
$K_{s}$ (AAT) & 21.89 $\pm$ 0.07 mag arcsec$^{-2}$ \\
$J-K_{s}$ (AAT) & 0.80 $\pm$ 0.09\\
$H-K_{s}$ (AAT) & 0.27 $\pm$ 0.08\\
 & \\
\multicolumn{2}{c}{{\it SW blob}} \\[2pt]
RA (2000) & 13$^h$05$^m$15$^s$.8 ($\pm$1$^{\prime\prime})$\\
Dec (2000) & --29$^{\circ}$34$^{\prime}$45$^{\prime\prime}$ ($\pm$1$^{\prime\prime})$\\
$g$ (APASS system) & 24.8 $\pm$ 0.1 mag arcsec$^{-2}$\\
$g-r$ (APASS system) & 0.93 $\pm$ 0.10\\
$K_{s}$ (AAT) & 21.69 $\pm$ 0.06 mag arcsec$^{-2}$\\
$J-K_{s}$ (AAT) & 0.67 $\pm$ 0.07\\
$H-K_{s}$ (AAT) & 0.18 $\pm$ 0.06\\ 
\hline
\end{tabular}
\end{center}
\end{table}

\begin{figure}
\centering
\includegraphics[angle=-90.,width=0.46\textwidth]{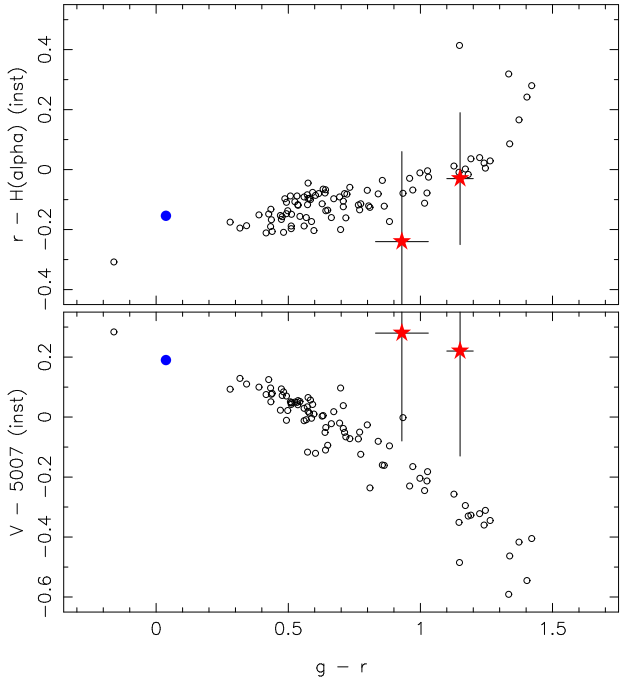}
\caption{Upper panel: the relationship between $(g-r)$ colour, on the APASS system, and ($r$--H$\alpha$).
The H$\alpha$ magnitudes are on the observed instrumental system.  Stars in the field of SMSS J1305--2931 are plotted 
as open circles, the central star is plotted as blue filled circle and the blobs are plotted as red star-symbols.  
Lower panel: as for the upper panel except that the relation is for $V-5007$.
\label{FS_phot_fig1}}
\end{figure}

Aperture photometry was next carried out on the ($r$, H$\alpha$) combined frame pair.  
The $r$ magnitudes were again transformed to the 
APASS system and used to calculate the ($r$--H$\alpha$) colours, where the H$\alpha$ magnitudes 
remain on the instrumental system but with a zero point chosen so that ($r$--H$\alpha$) $\approx$ 0 for 
stars with $(g-r)$ $\approx$ 1.0.  
The [($r$--H$\alpha$), $(g-r)$] two-colour diagram for the stars in the field with $g$ $\leq$ 19.3 is shown 
in the upper 
panel of Fig.\ \ref{FS_phot_fig1}.  The stars define a sensible sequence, and there is some indication that 
SMSS J1305--2931 shows a H$\alpha$ excess relative to the field star relation, though admittedly this is 
set by the single star in the field observed that is bluer than SMSS J1305--2931.  The presence of such 
an excess, however, is not unexpected given that the optical spectrum of SMSS J1305--2931 shows 
H$\alpha$ in emission (see \S \ref{Spectra}).

 
The derived $(r-H\alpha)$ colours for the blobs are plotted in the upper panel Fig.\ \ref{FS_phot_fig1} 
against $(g-r)$.  While uncertain at the 0.2 -- 0.3 mag level, there is no evidence of the presence 
of significant H$\alpha$ excess for the blobs.  This suggests that an optical spectrum 
of the blobs may not show substantial H$\alpha$ emission.  The errors in the colours come from 
varying the aperture locations by $\pm$1 pix at fixed aperture size and by varying the aperture size 
at a fixed center.   
Using the value of 15\AA\/ for the mean H$\alpha$ equivalent width (EW) in the spectrum of the central 
star (\S \ref{Spectra}, Fig.\ \ref{wifes_halpha_fig}) and its location approximately 0.1 mag above the 
field star locus in Fig.\ \ref{FS_phot_fig1}, the combined mean uncertainty in the blob $(r-H\alpha)$ 
colours (0.18 mag) then sets a loose upper limit of $\sim$25\AA\/ for the equivalent width of any 
H$\alpha$ emission from the blobs.  Integrated spectra of the blobs are required to place a tighter limit;
we note that the detectability of any emission line relative to the continuum in such an integrated 
spectrum will be strongly dependant on the emission line width.

We then conducted an essentially identical procedure for the ($V$, [O{\sc iii}]) frame pair, again using the 
APASS catalogue to calibrate the $V$ magnitudes.  The [$(V-$[O{\sc iii}]), $(g-r)$] 2-colour diagram for the 
stars is shown in the lower panel of Fig.\ \ref{FS_phot_fig1},  where we have set the zero point of the 
$(V-$[O{\sc iii}]) colours such that $(V-$[O{\sc iii}]) $\approx$ 0 at $(g-r)$ $\approx$ 0.6.  
Since the blobs are not strongly evident on the [O{\sc iii}] frames, care was taken to ensure that the 
aperture centre corresponded to the same position as for the $g$, $r$ and H$\alpha$ frames, and the same aperture 
size was used.  Similar precautions were applied for the $V$ frame.  The measurements on the [O{\sc iii}] frame 
did result in an apparent excess above sky, and the resulting $(V-$[O{\sc iii}]) colours are shown in the lower
panel of Fig.\ \ref{FS_phot_fig1}.  As for the ($r$--H$\alpha$) frames, the error bars for the ($V-$[O{\sc iii}]) 
colours represent the variations from changing the aperture size and position.  The $(V-$[O{\sc iii}]) colours 
appear to indicate an excess above the stellar sequence of a similar amount ($\sim$0.5 mag) for both blobs, perhaps indicating the presence of [O{\sc iii}] emission.  The excess, however, is less than 
1.5$\sigma$, and must 
be regarded as very uncertain until confirmed or refuted by better data.  {\it We therefore conclude that 
the optical radiation from the blobs is likely to be dominated by continuum rather than emission-line flux}.

\begin{figure}
\centering
\includegraphics[angle=-90.,width=0.46\textwidth]{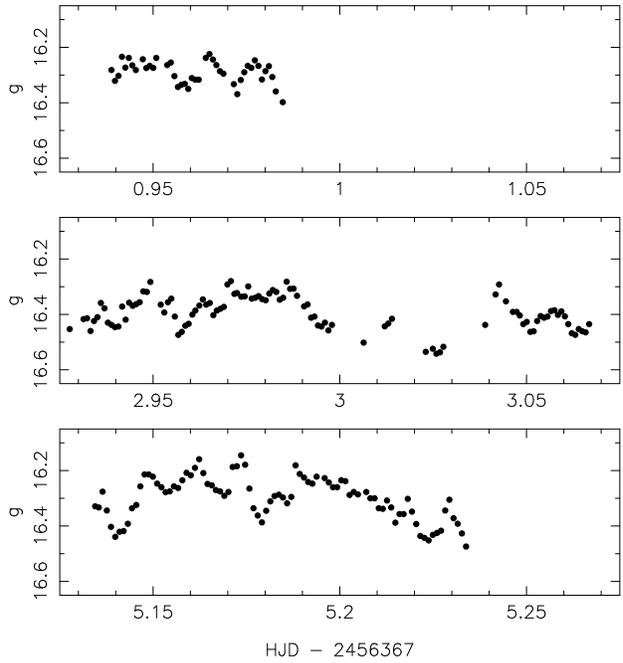}
\caption{Magnitude of the central star of the SMSS J1305--2931 system in the APASS $g$-band system as a 
function of heliocentric Julian Date, from repeated observations over three nights with the Faulkes-S telescope.  
The data indicate that the central star is variable by as much as $\sim$0.2 mag on timescales as short as 10--15 min; the uncertainty in each individual measurement is of order 0.02 mag.
\label{FS_phot_fig2}}
\end{figure}

Finally, the optical variability of the central star was investigated on the 2013 March $g$-band images via 
differential aperture photometry relative to a number of other stars of similar magnitude on the frames. 
The light curves for 2013 March 16, 18 and 20 are shown in Fig.\ \ref{FS_phot_fig2}.  These data show 
that the central star of the SMSS J1305--2931 system varies by
$\sim$0.2 mag on timescales as short as 10--15 min, with no obvious periodicity.  There is also a likely 
longer-term variation as the mean brightness of the star is $\sim$0.1 mag fainter on 2013 March 18 
compared to the other two nights.  The mean $g$ magnitude and standard deviations for the three nights 
are 16.29 and 0.04 mag, 16.40 and 0.06 mag, and 16.30 and 0.08 mag, respectively.  The mean 
magnitudes are all consistent with the value $g$ = 16.36 $\pm$ 0.06 given in the APASS catalogue.
In this process we also derived a $(g-r)$ colour of 0.04 for 
the central star on the APASS system, which is again consistent with the colour (0.01) 
given in the APASS catalogue.  

\subsubsection{AAT near-IR Imaging} \label{NIRphot}

Images of the SMSS J1305--2931 field were obtained in the $J$, $H$, and $K$ bands with the IRIS2 near-IR imager 
\citep{CGT04} on the AAT on 2014 January 27 via the AAO Service Observing program.  Conditions were 
photometric with seeing varying between 1$^{\prime\prime}$.7 and 2$^{\prime\prime}$.4.  Total integration
times were 30 min in $J$, 80 min in $H$, and 50 min in $K$, obtained as sets of dithered 10~s exposures.  The raw 
data frames were processed with the AAO in-house ORAC-DR data reduction software 
suite\footnote{http://www.aao.gov.au/science/instruments/current/IRIS2\\
/reduction}, which corrects for astrometric distortion of the camera prior to aligning and combining the images into 
the final mosaic.  

The zero points for the instrumental magnitudes measured on the combined frames were
 determined via a 
comparison with the magnitudes listed in the 2MASS point source catalogue \citep{SC06} for typically 12--14 stars on the frames.  
The resulting 1$\sigma$ uncertainty in the zero points is 0.01 mag for $J$ and 0.02 mag for $H$ and $K$\@.  
Application of these zero points then yields the $K$ magnitude and the $J-K$ and $H-K$ colours for the central star
listed in Table \ref{Tab1}.  
The $J$ and $H$ magnitudes are in good agreement with the values listed in the 2MASS catalogue, 
but the $K$ mag is 0.27 mag brighter.  Given that the 2MASS $K$ magnitude for the central star has a listed uncertainty of 0.27 mag, the difference is not significant.

\begin{figure}
\centering
\includegraphics[angle=0.,width=0.454\textwidth]{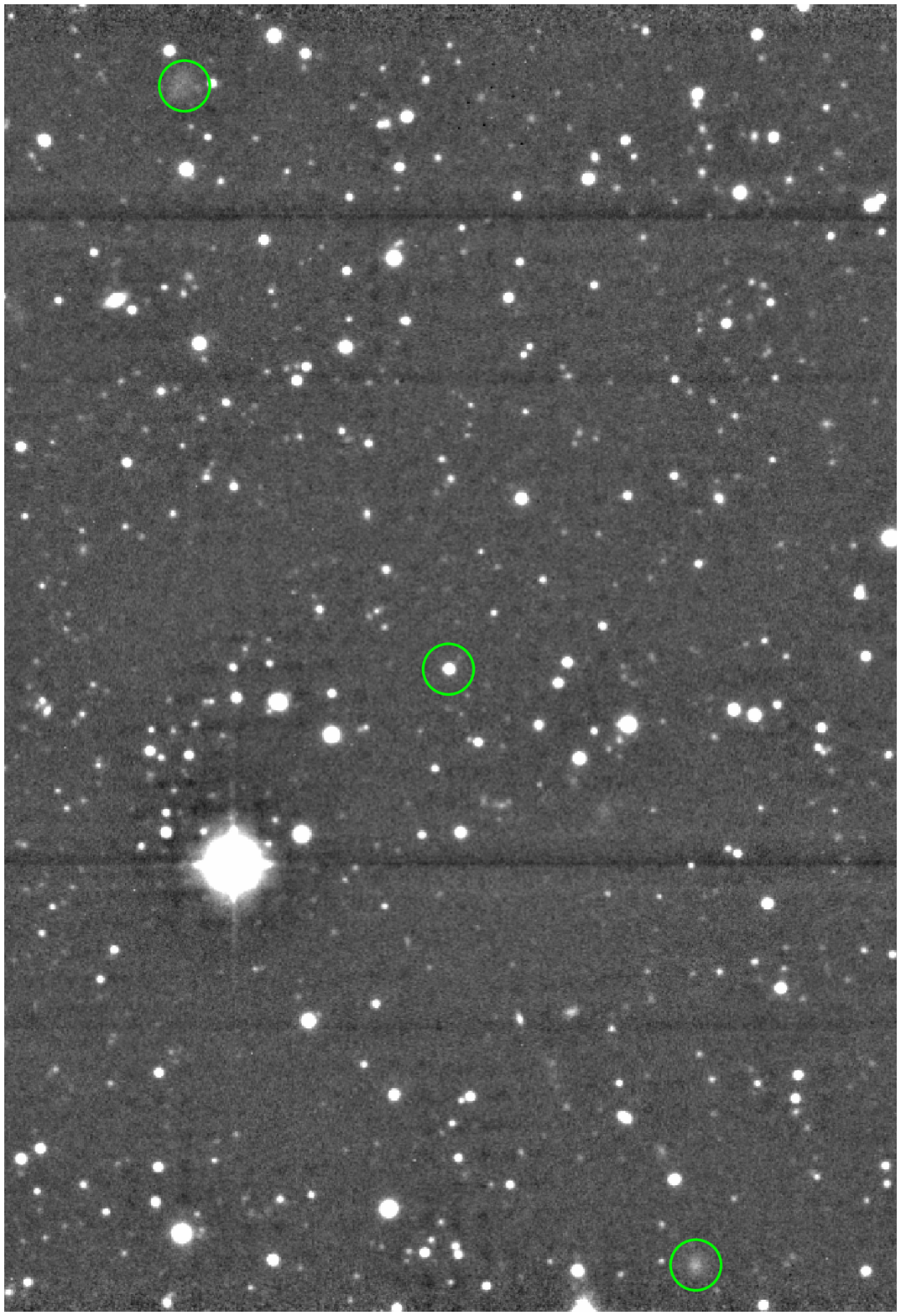}
\begin{minipage}{0.46\textwidth}
\centering
\includegraphics[angle=0.,width=0.494\textwidth]{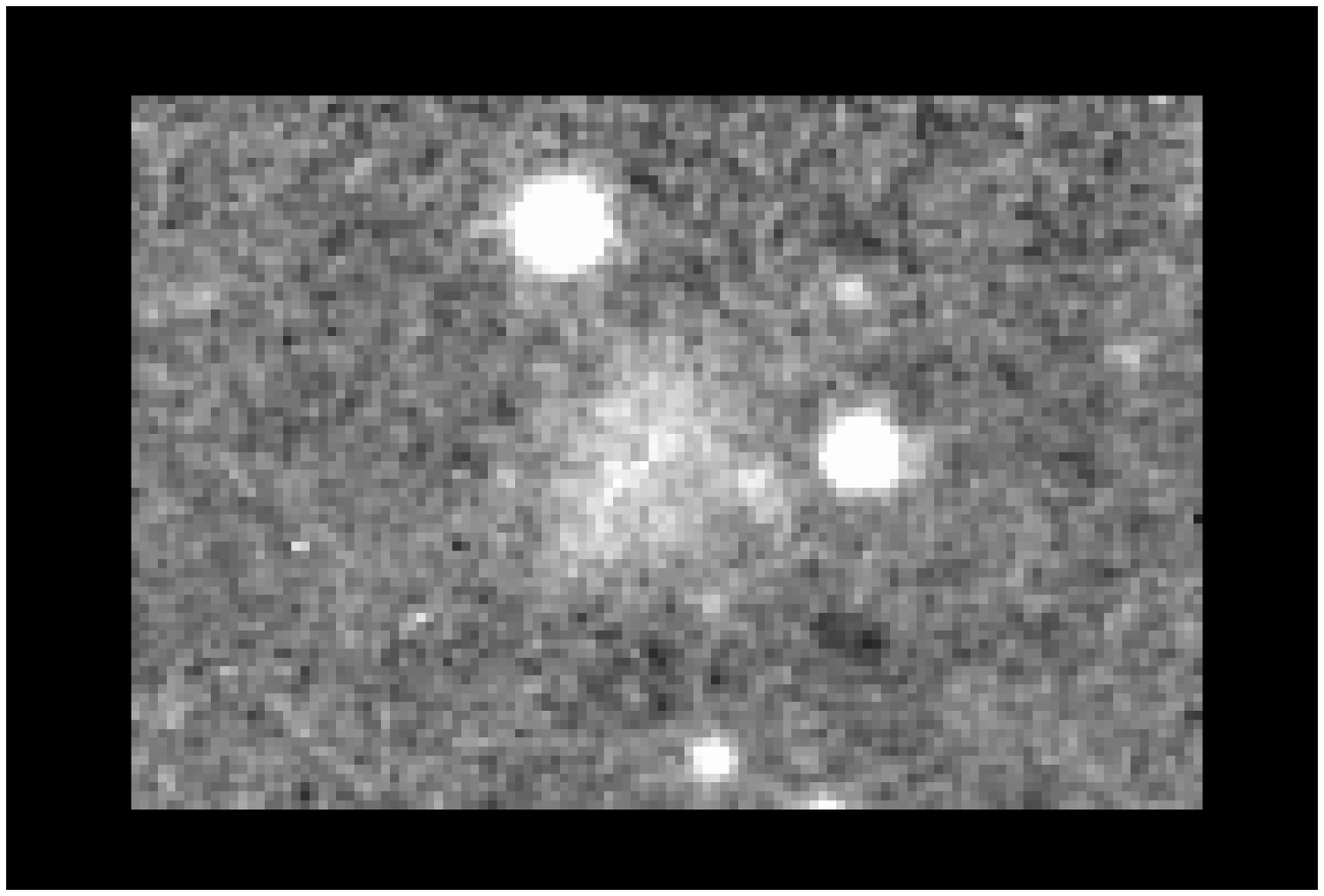}
\includegraphics[angle=0.,width=0.494\textwidth]{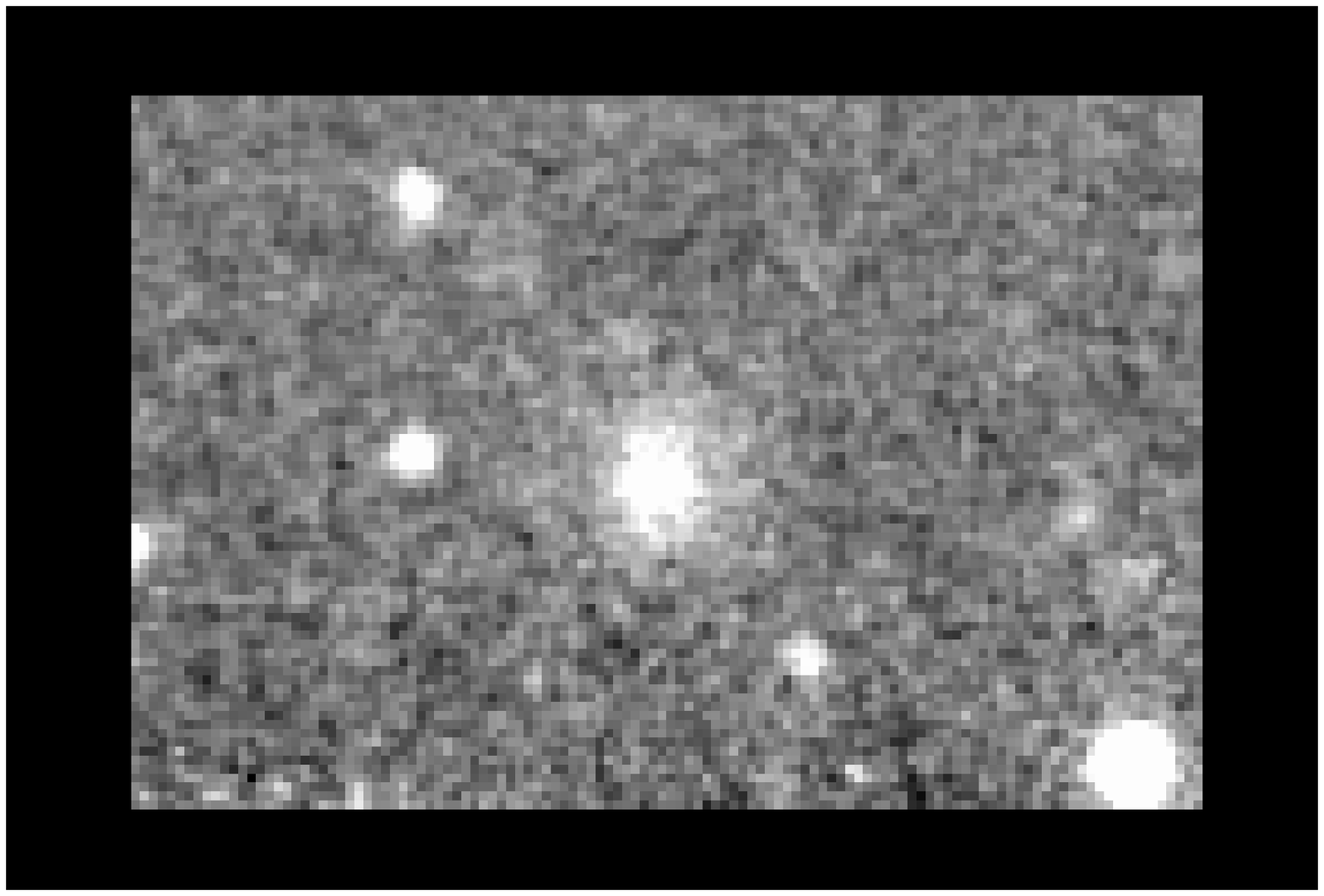}
\end{minipage}
\caption{{\it Upper panel}: A sub-section of the combined 80 min total integration $H$-band image of the SMSS J1305--2931 field obtained with IRIS2 
on the AAT on 2014 January 27.  The central star and the NE and SW blobs are circled.  The horizontal lines are artifacts arising from saturated stars.  
North is to the top and East to the left, and the circles have a diameter of 18$^{\prime\prime}$.
The field shown is 5$^\prime$.3 $\times$ 7$^\prime$.7 (EW $\times$ NS) in size.
{\it Lower panels}: Enlargements of the regions of the NE (left panel) and SW (right panel) blobs; 
the less centrally concentrated nature of the NE blob, compared to the SW blob, is evident.  Orientation is the same as the upper panel and the panels are $54^{\prime\prime} \times 36^{\prime\prime}$ in size.
\label{H_phot_fig}}
\end{figure}

The upper panel of Fig.\ \ref{H_phot_fig} shows the final $H$-band combined image for a
5$^\prime$.3 $\times$ 7$^\prime$.7 (EW $\times$ NS) sub-section of the full field: the NE and SW blobs are clearly evident, and the lower panels show enlargements of areas centered on each blob. The blobs are also readily visible on the $J$- and $K$-band images.  

To measure their magnitudes, we adopted the pixel-coordinates of the 
centres of the blobs on the $H$-band frame and then corrected them for the pixel offsets to the $J$ 
and $K$ frames determined from nearby stars.  Aperture magnitudes were then measured at these 
centres with an aperture of 8 pixel radius (3$^{\prime\prime}$.6), which encompasses most of the light 
from the blobs.  Application of the zero point values then yields the mean surface brightness and colours 
given Table \ref{Tab2};  as for the optical data it is clear that the two blobs again have similar colours and surface brightnesses.  

\subsection{Optical Spectroscopy} \label{Spectra}

Follow-up spectroscopy of the central star of the SMSS J1305--2931 system\footnote{The faint surface brightness of the blobs makes their spectroscopic observation very difficult with the telescopes to which we had ready access and it was not attempted.}
was obtained with the ANU 2.3m telescope on 2012 May 29, 2012 June 8--10 and 2012 June 12 with
the WiFeS double-beam integral field spectrograph \citep{MAD10}.  For the 
first four nights the B3000 and R3000 gratings were employed, which give coverage of 
essentially the entire optical window at $R$ $\approx$ 3000, while 
on the fifth night the R7000 grating was used in the red arm, yielding coverage in the vicinity 
of H$\alpha$ at $R$ 
$\approx$ 7000.  Integration times varied between 900~s and 2400~s with one exposure per night
except for the fifth night, when two 2000~s exposures were obtained that were subsequently combined.

Spectra were also obtained with the SOAR 4.1m telescope and the Goodman
High Throughput Spectrograph\footnote{http://www.soartelescope.org/observing/documentation/\\
goodman-high-throughput-spectrograph/goodman-manual/manual}
on 2012 July 6 and 2012 August 7 -- 9.  Spectra were obtained with a 1200 line grating covering the wavelength 
region $\lambda\lambda$4250--5500\AA\/ with $R$ $\sim$ 3000 and a
1.03$^{\prime\prime}$ slit.  On 2012 July 6 a single 1200 s exposures was obtained, while for 2012 August 10 -- 13 
three 1200 s exposures were obtained each night.  Both the WiFeS and the Goodman HTS spectra were continuum 
normalised prior to the analysis.

Initial inspection of the blue spectra shows a general similarity to the discovery spectrum (Fig.\ \ref{Sp_Fig1}), but 
closer inspection reveals that there are night-to-night variations in the strength of the emission lines relative to the 
continuum, and in their profile and location.  This applies also to the red spectra, which are dominated by 
H$\alpha$ in emission, but which also show He {\sc i} emission lines at $\lambda$5876, 6678 and 7065\AA.  
Na~D absorption is also present at a heliocentric velocity close to zero (the average heliocentric velocity
of the Na~D lines from 2012 June 8, 9 and 10 red spectra is 19 $\pm$ 10 km s$^{-1}$).  Intriguingly, using the relation
provided in \citet{MZ97}, the equivalent width of the Na D$_{1}$ line in the averaged red spectrum from the 2012 June 8, 9
and 10 observations implies E($B-V$) $\approx$ 0.3, or $\sim$0.2 mag in excess of the inferred line-of-sight
reddening.  There is no evidence for multiple components in the Na~D absorption, but given the resolution, velocity differences
in excess of 200 km s$^{-1}$ would be required to resolve separate components.  High-resolution spectra would be needed 
to distinguish any possible source component from that of the interstellar medium.

\begin{figure}
\centering
\includegraphics[angle=-90.,width=0.46\textwidth]{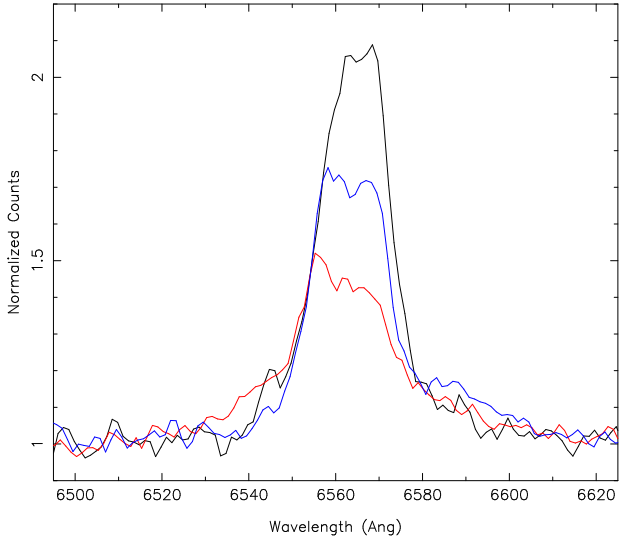}
\caption{Continuum normalized red spectra from the WiFeS observations on 2012 June 8 (black, highest peak 
counts), June 9 (red, lowest peak counts) and June 10 (blue, intermediate peak counts) in the vicinity
of the H$\alpha$ line, illustrating the night-to-night changes in the line profile, relative strength and position.
\label{wifes_halpha_fig}}
\end{figure}

In Fig.\ \ref{wifes_halpha_fig} we show the continuum-normalised H$\alpha$ profiles from the WiFeS 
observations on 2012 June 8, 9, and 10.  
These profiles are typical of those seen in the red spectra.  Relative to the continuum, 
the H$\alpha$ emission flux dropped by $\sim$50\% from June 8 to June 9, and then
increased again on June 10.  
The line profiles can be fit by single Gaussian profiles in some instances (e.g.\ May 29, June 8, June 9) 
but in other cases multiple components are present.  For example, the line profiles for June 10 and 
June 12 are convincingly fit by two components with similar FWHM values.  The derived
H$\alpha$ line widths and velocities, and their uncertainties, are summarized in Table \ref{Table3}.


\begin{figure}
\centering
\includegraphics[angle=-90.,width=0.46\textwidth]{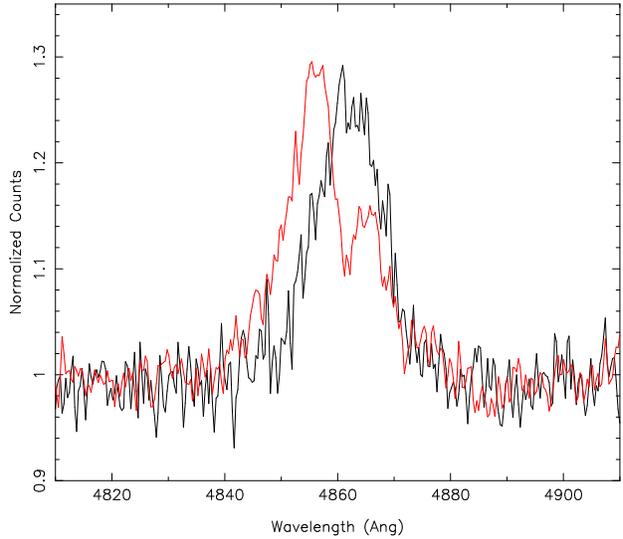}
\caption{Continuum normalized blue spectra from the SOAR observations obtained on 2012 August 7 (black; single 
peak at longer wavelengths) and 2012 August 9 (red; dual peaks) in the vicinity
of the H$\beta$ line, illustrating the changes in the line profile and position.
\label{soar_bluesp_fig}}
\end{figure}

The H$\beta$ profiles extracted from the WiFeS blue spectra are generally similar to the H$\alpha$ profiles: single-peaked on May 29 and June 8--9, and double-peaked on June 10 and 12.
For the three cases where a single Gaussian line profile fit is preferred, the $H\beta$ velocities are 
$\sim250\pm50$ km s$^{-1}$more negative than the corresponding contemporaneous H$\alpha$ velocities, although there is no obvious 
difference in the line widths.  On the other hand, at the $\sim$50 km s$^{-1}$ level, the velocities of
the corresponding components agree when the line profiles are double: e.g., for June 10 the H$\beta$ 
components are at --285 and +210 km s$^{-1}$, while the H$\alpha$ components are at
--223 and +240 km s$^{-1}$.  The respective line widths are again broadly comparable.  
The H$\beta$ profiles from the Goodman HTS spectra have similar properties to those seen in the 
WiFeS blue spectra obtained two months earlier, changing between single-peaked and double-peaked, 
sometimes over successive nights.  Fig.\ \ref{soar_bluesp_fig} shows the Goodman HTS 
spectra in the vicinity of H$\beta$ for 2012 August 7 (single component) and August 9 (two 
components).   The H$\beta$ velocities and line widths derived from the SOAR spectra are also
listed in Table \ref{Table3}.  
We note that the average EW of the H$\beta$ emission across all our blue 
spectroscopic observations is $\langle$EW(H$\beta$)$\rangle$ $\approx$ 4.4 \AA, while that 
for He {\sc ii} $\lambda$4686\AA\/ is
$\langle$EW(He {\sc ii})$\rangle$ $\approx$ 4 \AA\/.


In summary, the spectroscopic observations indicate substantial variations in the Balmer
emission line strengths, profiles, 
and velocities on timescales at least as short as night-to-night.  The velocity changes have scales of 
100s km s$^{-1}$, and the line widths vary between $\sim$300 and $\sim$1000 km s$^{-1}$.  These properties point to emission from an accretion disk, or more generally, from warm gas moving in the vicinity of an accreting object.  More detailed dedicated spectroscopic observations are needed to 
investigate whether these emission line variations occur on timescales as short as those seen in the 
broadband photometry, and whether any periodic motions are discernible.

\begin{table}
\caption{Examples of the velocity structure of the Balmer emission lines in the spectrum of the 
central source.}
\label{Table3}
\begin{center}
\begin{tabular}{lcc}
\hline
Date &  $\gamma^{a}$ & FWHM \\
   & (km s$^{-1}$) & (km s$^{-1}$)\\
\hline
\multicolumn{3}{c}{H$\alpha$ from ANU 2.3m} \\[2pt]
2012 May 29   & $-130 \pm 15$ &  $750\pm 100$  \\
2012 June 8  &  $+50\pm 15$ & $800 \pm 100$   \\
2012 June 9    &  $-40 \pm 15$ & $1100 \pm 100$  \\
2012 June 10$^b$  & $(-20 \pm 15)$  & $(860 \pm 100)$   \\
      &  $-223 \pm 15$ &  $450 \pm 100$  \\
    &  $+240 \pm 15$ &  $450 \pm 100$     \\
2012 June 12$^b$   &  $(+15 \pm 15)$ & $(870 \pm 100)$  \\
   & $-300 \pm 15$  &  $400 \pm 100$ \\
    & $+170 \pm 15$&  $400 \pm 100$\\
\hline
\multicolumn{3}{c}{H$\beta$ from SOAR 4.1m}  \\[2pt]
2012 July 6    & $-240 \pm 15$ &  $680 \pm 100$\\
    &   $+265 \pm 15$ &   $330 \pm 100$   \\
2012 August 7   & $+15 \pm 15$ &  $850 \pm 100$ \\
2012 August 8  & $-230 \pm 15$ &  $650 \pm 100$  \\
    &   $+240 \pm 15$  &   $250 \pm 100$   \\
2012 August 9   & $-380 \pm 15$ &  $600 \pm 100$ \\
     &   $+270 \pm 15$  &   $330 \pm 100$  \\
\hline
\multicolumn{3}{l}{$^a$ Heliocentric systemic velocity.}\\
\multicolumn{3}{l}{$^b$ Two-Gaussian fit preferred on those dates.}\\
\end{tabular}
\end{center}
\end{table}
 
\subsection{UV and X-ray Observations}  \label{X-ray}

\subsubsection{Swift/UVOT observations}

The 30-cm UltraViolet and Optical Telescope (UVOT) \citep{Ro05} is mounted onboard the {\it Swift} 
spacecraft \citep{Ge04} and is co-aligned with the 0.3 -- 10 keV X-Ray Telescope \citep[XRT;][]{Bu05}. 
UVOT has a 17$^{\prime}\times$17$^{\prime}$ field-of-view and is equipped with six broad-band filters, 
three in the UV ({\it uvw2, uvm2, uvw1}) covering the wavelength range 1600--3000\AA, and three in the 
optical ({\it u, b, v}), which correspond approximately to standard $U$, $B$ and $V$ \citep{Po08}.  
The SMSS J1305-2931 system was observed with UVOT during the 
period 2012 July 6 -- 2012 Sept 2 coincident with the XRT observations described in \S \ref{XRT} below.
It was observed in all six UVOT filters with one integration in each filter per orbit.  A total of 16 or 17 
observations were obtained for each filter and the data frames were analyzed as described in 
\citet{Po08}.  

There was no evidence for the presence of the blobs in any of the images,  but the central star was readily detected in all exposures.  The mean magnitudes (VEGAMAG system) are listed in Table \ref{Tab1}. The individual magnitudes 
confirm the variability of the central source on timescales as short as the $\sim$95 min interval
between successive spacecraft orbits.  The intrinsic standard deviations of the magnitudes are of order 
$\sim$0.1 -- 0.15 mag, while the range in the observed magnitudes across the $\sim$60 day span is 
$\sim$0.4 mag for all six filters.  There is also a hint that the
star is slightly bluer when brighter; confirmation of this effect is made difficult by the relatively large errors in the colours (typically 0.15 mag).

\subsubsection{Swift/XRT and XMM-Newton observations} \label{XRT}

The field containing SMSS 1305--2931 had previously been observed in X-rays only with with the 
{\it ROSAT} Position Sensitive Proportional Counters (PSPC) detector as part of the {\it ROSAT} 
All-Sky survey \citep{Voges99}.  No X-rays were detected at the source position in a 348 s observation,
and the 3$\sigma$ upper limit on the count rate is estimated as 0.017 counts s$^{-1}$, corresponding 
to an upper limit of $\sim$6 $\times$ 10$^{-13}$ erg cm$^{-2}$ s$^{-1}$ in the 0.3--10 keV energy range, 
assuming a Galactic absorbing column of 7.15 $\times$ 10$^{20}$ atoms cm$^{-2}$ \citep{Ka05}
and a power-law spectrum with a photon index $\Gamma = 1.7$.

Subsequent to our optical discovery, we were awarded a Target of Opportunity monitoring
program with {\it Swift}/XRT\@.  The field of SMSS 1305--2931 was observed for intervals of $\sim$2000--3000~s approximately weekly over a three month period, and a total of 16.5~ks 
on-source integration was ultimately obtained.  An X-ray counterpart to SMSS J1305--2931 was 
detected at a 3.9$\sigma$ level above the background.  There was, however, no indication of any 
X-ray emission at the location of either of the blobs.

No significant variability was seen in the XRT observations of the central source, but the net counts
from each observation are inadequate to draw firm conclusions.
The average net count rate 
is approximately 2 $\times$ 10$^{-3}$ counts s$^{-1}$ in the 0.3 -- 10 keV range, 
corresponding to an absorbed flux of $\sim$7 $\times$ 10$^{-14}$ erg cm$^{-2}$ s$^{-1}$ for a 
power-law photon index $\Gamma$ = 2, or $\sim$1.4 $\times$ 10$^{-13}$ for $\Gamma$ = 1.  
The 3$\sigma$ upper limits for any emission from the blobs, which lie off-axis, are 
1 $\times$ 10$^{-3}$ counts s$^{-1}$ for the NE feature and 4 $\times$ 10$^{-3}$ counts s$^{-1}$ 
for the SW feature.  For a power-law index $\Gamma$ of 1.7 the corresponding flux limits are 4 $\times$ 10$^{-14}$ and 2 $\times$ 10$^{-13}$ erg cm$^{-2}$ s$^{-1}$, respectively.

Following our {\it Swift}/XRT detection, we obtained a deeper observation of the field with
{\it XMM-Newton} (OBSID: 0723380101, PI: Farrell): a 70~ks exposure on 2014 January 12, with the 
European Photon Imaging Camera (EPIC) \citep{struder01,turner01}  as the prime instrument.
The data files were 
processed with the Science Analysis System ({\small{SAS}}) version 14.0.0 (xmmsas\_20141104).   
Sub-intervals of the observation with high particle background were filtered out and time intervals of 
38ks for EPIC-pn and 62 ks for EPIC-MOS retained.    
As before, the central object was clearly detected but there is no indication of any X-ray 
emission from either of the blobs.  

The {\it XMM-Newton} data provided enough counts for spectral analysis. 
We used the {\small SAS} task {\it xmmselect} to extract spectra and background files for the pn and MOS data, using a circular source region of 20\arcsec\ radius, and a local background region three times as large (avoiding chip gaps). We used the standard flagging criteria \verb|#XMMEA_EM| for the MOS data, and \verb|FLAG=0| \&\& \verb|#XMMEA_EP| for the pn data. We built response and ancillary response files with the {\small SAS} tasks {\it rmfgen} and {\it arfgen}. Finally, to increase the signal-to-noise ratio, we combined the pn, MOS1 and MOS2 spectra with {\it epicspeccombine}\footnote{https://www.cosmos.esa.int/web/xmm-newton/sas-thread-epic-merging}, to create a weighted-average EPIC spectrum. We grouped the combined spectrum to a minimum of 25 counts per bin, for Gaussian statistics. As a consistency check, we later repeated our modelling by fitting the individual pn and MOS spectra simultaneously, and obtained similar results to those described below.

\subsubsection{Modelling the X-Ray spectrum}

In order to provide additional information to assist with the interpretation of the source, we modelled the {\it XMM-Newton}/EPIC spectrum with {\footnotesize{XSPEC}} version 12.9.1 \citep{arnaud96}.  We note first that simple models, such as an absorbed single-temperature blackbody, a power-law, or a single-temperature thermal plasma, do not provide satisfactory fits. The lack of satisfactory fits is also true for standard two-component models that are often applied to accreting compact objects ({\it e.g.}, blackbody plus power-law or disk-blackbody plus power-law). These models, however, do provide the physical insight that the X-ray spectrum appears to have both a hard and a soft component. A good fit to the combined EPIC spectrum is obtained with models employing two optically-thin-plasma thermal components, with different absorption column densities for the soft and hard components. 

In the simplest version of this model, we took a single-temperature plasma component ({\it mekal} in {\small {XSPEC}}) for the soft X-ray emission, and a single-temperature component for the hard X-ray emission, with two different accretion columns ({\it i.e.}, assuming that the two components come from different spatial regions in the system). The resulting best-fitting parameters (Table \ref{Tab_xray}) are a temperature $kT_{\rm s} \approx 0.2$ keV for the soft component (absorbed by a column density $n_{\rm H,s} \approx 0.9 \times 10^{22}$ cm$^{-2}$) and $kT_{\rm h} \sim 10$ keV for the hard component (absorbed by a column density $n_{\rm H,h} \approx 4 \times 10^{22}$ cm$^{-2}$).
The observed flux in the 0.3--2 keV band is $f_{0.3-2} \approx 7 \times 10^{-15}$ erg cm$^{-2}$ s$^{-1}$ (heavily affected by absorption), and $f_{0.3-10} \approx 1.1 \times 10^{-13}$ erg cm$^{-2}$ s$^{-1}$ in the 2--10 keV band.
The unabsorbed luminosity is well constrained for the hard band, at $L_{2-10} \approx 7 \times 10^{31} \times d^2_{2{\rm kpc}}$ erg s$^{-1}$ (where $d_{2{\rm kpc}}$ is the distance to the source in units of 2~kpc), but is very uncertain for the soft band, due to the low temperature of the emitting plasma and the high column density; we estimate $L_{0.3-2} \approx {\rm {a~few}} \times 10^{32} \times d^2_{2{\rm kpc}}$ erg s$^{-1}$ (Table \ref{Tab_xray}).
However, the two-temperature approximation, while formally a good fit ($\chi^2_{\nu} = 44.7/45 = 0.99$), is a crude physical model, especially for the soft component; it may lead to an over-estimate of the column density and therefore of the intrinsic luminosity for the soft component.

\begin{figure}
\centering
\includegraphics[angle=-90.,width=0.46\textwidth]{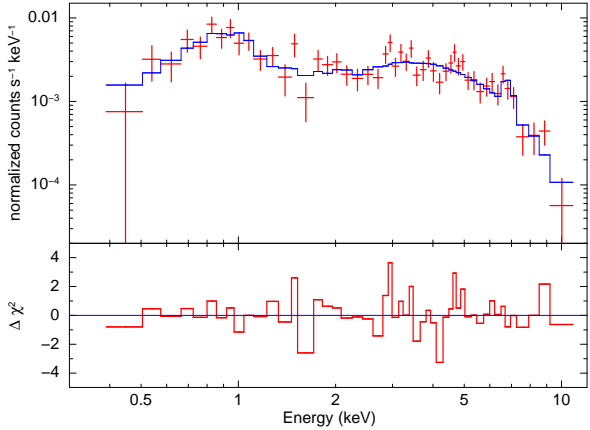}
\caption{The {\it XMM-Newton}/EPIC spectrum of the SMSS J1305--2931 X-ray counterpart compared with the best-fit multi-temperature thermal-plasma model.  The observed points are shown as red plus-signs with error bars, while the model fit is shown in blue.  The lower panel shows the $\Delta\chi^{2}$ difference between the model and the data.  The parameters and physical interpretation of the model are discussed in section \ref{XRT}.
\label{xray_fig}}
\end{figure}

We refined our model by assuming a multi-temperature distribution ({\it cemekl} in {\small {XSPEC}}) for the softer emitter (Table \ref{Tab_xray}). This model provides an equally good fit ($\chi^2_{\nu} = 42.7/45 = 0.95$). The maximum temperature of the {\it cemekl} distribution has to be $>$1.3 keV; the temperature of the hotter component is again $kT_{\rm h} \sim 10$ keV. The unabsorbed luminosities in this model are $L_{0.3-2} \approx 4 \times 10^{31} \times d^2_{2{\rm kpc}}$ erg s$^{-1}$ and $L_{2-10} \approx 8 \times 10^{31} \times d^2_{2{\rm kpc}}$ erg s$^{-1}$ (Table \ref{Tab_xray}), for a total X-ray luminosity $L_{0.3-10} = (1.2 \pm 0.2) \times 10^{32} \times d^2_{2{\rm kpc}}$ erg s$^{-1}$.
The data points, model fit, and $\chi^2$ residuals for this model are shown in Fig.\ \ref{xray_fig}.

Finally, we tried fitting the spectrum with only one multi-temperature distribution of plasma ({\it cemekl}), absorbed by a moderately low column density plus a higher column density component with covering fraction $< 1$ ({\it pcphabs} $\times$ {\it TBabs} $\times$ {\it cemekl}). This could be the situation, for example, if the absorbing material is clumpy, or if some of the hot emitting gas is located above the densest part of the absorbing material. We obtain $\chi^2_{\nu} = 42.0/46 = 0.91$, statistically equivalent to the two-component emission model.  The baseline column density is
$n_{\rm H,1} \approx 0.23 \times 10^{22}$ cm$^{-2}$;
the partial absorbing component has
$n_{\rm H,2} \approx 6 \times 10^{22}$ cm$^{-2}$, with covering fraction
$f_{\rm cvr} = 0.94 \pm 0.02$.
The thermal plasma requires a maximum temperature $T_{\rm max} > 24$ keV (the precise value is unconstrained by the data).
The unabsorbed X-ray luminosity is 
$L_{(0.3-10)} = (2.1^{+0.8}_{-0.6}) \times 10^{32} \times d^2_{2{\rm kpc}}$ erg s$^{-1}$ 
(Table \ref{Tab_xray}.
The fit residuals look essentially identical to those for the previous model shown in 
Fig.\ \ref{xray_fig}; the factor-of-two difference in the intrinsic luminosity between the two models is due to the different way of accounting for the absorption of the soft X-ray photons.

\begin{table}
\caption{Main X-ray spectral parameters}
\label{Tab_xray}
\begin{center}
\begin{tabular}{lr}
\hline
Parameter &  Value \\
\hline
\multicolumn{2}{c}{{\it TBabs}$_{\rm s}$ $\times$ {\it mekal}$_{\rm s}$ 
$+$ {\it TBabs}$_{\rm h}$ $\times$ {\it mekal}$_{\rm h}$} \\[5pt]
$n_{\rm H,s}$  &  $0.88^{+0.14}_{-0.20} \times 10^{22}$ cm$^{-2}$\\[4pt]
$n_{\rm H,h}$  &  $4.0^{+0.8}_{-0.7} \times 10^{22}$ cm$^{-2}$\\[4pt]
$kT_{\rm s}$  &  $0.19^{+0.08}_{-0.04}$ keV\\[4pt]
$kT_{\rm h}$  &  $12^{+36}_{-4}$ keV\\[4pt]
EM$_{\rm s}$  &  $16^{+35}_{-13} \times 10^{54} \times d_{2{\rm kpc}}^2$ cm$^{-3}$\\[4pt]
EM$_{\rm h}$  &  $4.8^{+0.7}_{-0.6} \times 10^{54} \times d_{2{\rm kpc}}^2$ cm$^{-3}$\\[4pt]
$\chi^2_{\nu}$  &  $0.99 (44.7/45)$ \\[4pt]
$f_{0.3-2}$   &  $0.7^{+0.1}_{-0.1} \times 10^{-14}$ erg cm$^{-2}$ s$^{-1}$\\[4pt]
$f_{2-10}$   &  $11^{+1}_{-1} \times 10^{-14}$ erg cm$^{-2}$ s$^{-1}$\\[4pt]
$L_{0.3-2}$   &  $3.3^{+5.3}_{-2.3} \times 10^{32} \times d_{2{\rm kpc}}^2$ erg s$^{-1}$\\[4pt]
$L_{2-10}$   &  $7.1^{+0.4}_{-0.4} \times 10^{31} \times d_{2{\rm kpc}}^2$ erg s$^{-1}$\\[4pt]
\hline
\multicolumn{2}{c}{{\it TBabs}$_{\rm s}$ $\times$ {\it cemekl}$_{\rm s}$ 
$+$ {\it TBabs}$_{\rm h}$ $\times$ {\it mekal}$_{\rm h}$}  \\[5pt]
$n_{\rm H,s}$  &  $0.23^{+0.17}_{-0.13} \times 10^{22}$ cm$^{-2}$\\[4pt]
$n_{\rm H,h}$  &  $5.3^{+1.6}_{-1.3} \times 10^{22}$ cm$^{-2}$\\[4pt]
$\alpha_{\rm cemekl}$  & [0.01] \\[4pt]
$kT_{\rm max}$  &  $>1.3$ keV\\[4pt]
$kT_{\rm h}$  &  $12^{+24}_{-4}$ keV\\[4pt]
EM$_{\rm s}$  &  $0.24^{+0.46}_{-0.07} \times 10^{54} \times d_{2{\rm kpc}}^2$ cm$^{-3}$\\[4pt]
EM$_{\rm h}$  &  $4.9^{+0.8}_{-0.7} \times 10^{54} \times d_{2{\rm kpc}}^2$ cm$^{-3}$\\[4pt]
$\chi^2_{\nu}$  &  $0.95 (42.7/45)$ \\[4pt]
$f_{0.3-2}$   &  $0.8^{+0.1}_{-0.1} \times 10^{-14}$ erg cm$^{-2}$ s$^{-1}$\\[4pt]
$f_{2-10}$   &  $11^{+1}_{-1} \times 10^{-14}$ erg cm$^{-2}$ s$^{-1}$\\[4pt]
$L_{0.3-2}$   &  $4.1^{+0.8}_{-0.5} \times 10^{31} \times d_{2{\rm kpc}}^2$ erg s$^{-1}$\\[4pt]
$L_{2-10}$   &  $7.6^{+0.4}_{-0.4} \times 10^{31} \times d_{2{\rm kpc}}^2$ erg s$^{-1}$\\[4pt]
\hline
\multicolumn{2}{c}{{\it pcphabs} $\times$ {\it TBabs} $\times$ {\it cemekl}}  \\[2pt]
$n_{\rm H,1}$  &  $0.23^{+0.09}_{-0.10} \times 10^{22}$ cm$^{-2}$\\[4pt]
$f_{\rm {cvr}}$  & $0.94^{+0.2}_{-0.2}$ \\[4pt]
$n_{\rm H,2}$  &  $6.1^{+1.4}_{-1.1} \times 10^{22}$ cm$^{-2}$\\[4pt]
$\alpha_{\rm cemekl}$  & [0.01] \\[4pt]
$kT_{\rm max}$  & $> 24$ keV \\[4pt]
EM  &  $3.8^{+0.3}_{-0.4} \times 10^{54} \times d_{2{\rm kpc}}^2$ cm$^{-3}$\\[4pt]
$\chi^2_{\nu}$  &  $0.91 (42.0/46)$ \\[4pt]
$f_{0.3-2}$   &  $0.8^{+0.1}_{-0.1} \times 10^{-14}$ erg cm$^{-2}$ s$^{-1}$\\[4pt]
$f_{2-10}$   &  $11^{+1}_{-1} \times 10^{-14}$ erg cm$^{-2}$ s$^{-1}$\\[4pt]
$L_{0.3-2}$   &  $14^{+7}_{-2} \times 10^{31} \times d_{2{\rm kpc}}^2$ erg s$^{-1}$\\[4pt]
$L_{2-10}$   &  $7.6^{+0.4}_{-0.4} \times 10^{31} \times d_{2{\rm kpc}}^2$ erg s$^{-1}$\\[4pt]
\hline
\end{tabular}
\end{center}
\end{table}


\section{The Nature of the Central Object} \label{sect4}


The broad emission-line profiles in the optical spectrum, together with the variability of both the broadband continuum and the emission lines, strongly suggest that the dominant source of the radiation from the central object is a hot accretion disk around a compact object that is accreting mass from a companion star in a binary system.  The most likely candidates for the nature of the system are then a high-mass X-ray binary (HMXB), a low-mass X-ray binary (LMXB), or some sort of cataclysmic variable (CV).  We now discuss each of these possibilities.

We can rule out the HMXB classification straightaway.  In HMXB systems, by definition, the companion star is typically an O- or B-type star with a mass exceeding $\sim$8--10 
M$_{\sun}$, and as a result, the locations of HMXB systems are restricted to the Galactic plane \citep[e.g.][]{Chaty11}.  SMSS J1305--2931, however, lies far from the plane at a Galactic latitude $b$ of 33.25 deg
or a $z$-height of 1.1 $\times$ $d_{2{\rm kpc}}$ kpc, far away from star-forming regions.  Further, formation of an accretion disk in a HMXB system is not common \citep{Chaty11}, and in addition, there
is no evidence in the observed spectrum supporting the existence of a luminous donor-star, as would be expected.  Consequently, a HMXB interpretation for SMSS J1305--2931 is not viable.

LMXBs, on the other hand, are systems in which a low-mass star, typically a main sequence star of mass $\sim$1 M$_{\sun}$ or less, is losing mass via Roche-lobe overflow to a compact companion that is either a neutron star or a black-hole \citep[e.g.][]{TvdH10}.  LMXBs are generally associated with old stellar populations: they are found predominantly towards the Galactic Bulge, which is not inconsistent with the location of SMSS J1305--2931 on the sky, and in Galactic globular clusters.  The optical spectrum of an (active) LMXB is dominated by emission from the hot accretion disk surrounding the compact companion, again as seen for SMSS J1305--2931.   Optical line emission, from the X-ray irradiated surface of the accretion disk and/or donor star, in both neutron star and black hole LMXBs
\citep[e.g.][]{RS99,Casares06}, is also similar to that seen in the spectrum of SMSS J1305--2931.

However, there is one fundamental difference between the properties of LMXB systems and those of SMSS J1305--2931 that rules out the classification of SMSS J1305--2931 as an LMXB: the ratio of the X-ray luminosity to the optical luminosity.  For LMXBs $L_{\rm X} \sim 500-1000 L_{\rm opt}$ \citep[e.g.][]{vPM95,TvdH10}, with the exception of ``accretion-disk-corona'' (ADC) sources, seen at 
high inclination, for which  $L_{\rm X} \sim 20 L_{\rm opt}$ \citep[e.g.][]{Casares03}.  However,
such ratios are definitely ruled out for SMSS J1305--2931; instead $L_{\rm X} \ll L_{\rm opt}$.  Specifically, in \S\ref{XRT} we showed that the observed X-ray flux from SMSS J1305--2931 in the 0.3 to 10 keV range is $f_{0.3-10} = (1.2 \pm 0.1) \times 10^{-13}$ erg cm$^{-2}$ s$^{-1}$, whereas in the band-pass of the $g$-filter, the observed flux is of order 2 $\times 10^{-12}$ erg cm$^{-2}$ s$^{-1}$.  This indicates $f_{opt}$ is at least 20$\times$ the X-ray flux $f_{0.3-10}$; SMSS J1305--2931 is therefore unlikely to be an LMXB.

The remaining possibility for the central source of the SMSS J1305--2931 system is that it is a 
CV, in which a white dwarf is accreting matter from a low-mass main sequence or slightly evolved companion star via Roche-lobe overflow \citep[e.g.][]{warner95,hellier01}.  In non-magnetic CVs, which are the majority, the accretion process occurs via an extended accretion disk that surrounds the white dwarf.  The dominant sources of radiation are the disk itself and the boundary
layer between the innermost radius of the disk and the surface of the white dwarf.  For CV systems in which the white dwarf is strongly magnetic, the binary is tidally locked and the accretion is directly on to the white dwarf surface at the magnetic poles.  In these cases, known as polars (class prototype AM Her), there is an accretion column rather than an accretion disk.  Intermediate-polars (IPs, class prototype DQ Her) occur when the magnetic field is not sufficiently strong to cause tidal-locking -- there is an accretion disk, but the inner regions are truncated by the magneto-sphere of the white dwarf.  In all cases instability in the mass transfer and/or accretion processes leads to variability in the system.  Such variability is most extreme in the class of CVs known as dwarf novae, which show recurrent changes in brightness of up to 5 magnitudes \citep{warner95,coppejans16a}.  The changes are interpreted as a thermal limit-cycle instability in the accretion disk \citep[e.g.][]{Smak1983, Cannizzo1993, Osaki1996, Lasota2001}; the disk cycles between an ionized (high-luminosity) state and a neutral (low-luminosity) state, usually referred to as the upper and lower branches of the ``S-curve".  Finally, nova-like CVs are systems persistently in the high-luminosity state, probably because the accretion rate is high enough to maintain the disk on the upper branch of the S-curve \citep{Smak1983,warner95,balman14}.

We can immediately rule out a polar-CV classification for SMSS~J1305--2931 on the basis of the width 
of the emission lines in the optical spectrum --- polars present narrow spectral lines because of the 
absence of an accretion disk \citep[e.g.][]{Oliveira17}.  The lack of absorption lines in the optical 
spectrum, the presence of high-ionization lines (e.g.\ He {\sc{ii}} $\lambda$4686\AA), and the blue 
colours all point to SMSS J1305--2931 being in an accretion-dominated phase, suggesting a 
classification as an IP, or a dwarf nova, or a nova-like CV\@.  IPs are characterised by H$\beta$ 
equivalent widths in excess of 20\AA\/ \citep[e.g.][]{Oliveira17}, while the average value of $\langle
$EW({\rm H}$\beta$)$\rangle$ measured for SMSS J1305--2931 is $\sim$4.4 \AA\/  (\S \ref{Spectra}), 
ruling out an IP classification.   The value of $\langle$EW({\rm H}$\beta$)$\rangle$ is, however, typical 
of dwarf novae or nova-like CVs.


To further constrain the type of CV, we used the empirical relations \citep{Patterson1984, PR85a, PR85b, Hertz1990} between the observed soft X-ray flux (and observed X-ray/optical flux ratio) and the absolute visual magnitude of the system.
In those relations, the X-ray flux is usually taken in the 0.2--4.0 keV band for historical reasons, because early high-energy surveys of CVs were done with the {\it Einstein Observatory}. In SMSS J1305$-$2931, the observed 0.2--4.0 keV flux is $\approx 3 \times 10^{-14}$ erg cm$^{-2}$ s$^{-1}$ and the absorption-corrected flux is $\approx 1.5-3.5 \times 10^{-13}$ erg cm$^{-2}$ s$^{-1}$ depending on the details of the adopted model.  Optical fluxes are approximated by $\log F_{\rm opt} = -0.4V -5.44 \approx -12.0$ \citep{Schwope02}. With this definition, the observed $\log(F_{\rm X}/F_{\rm opt}) \approx -1.5$, while the absorption-corrected ratio is between
$-0.9$ and $-0.5$ depending on the choice of model. This range of flux ratios is typical of CVs with an accretion rate $10^{17}$ g s$^{-1}$ $\la \dot{M} \la $ a few $10^{18}$ g s$^{-1}$, assuming that the optical continuum comes from a standard accretion disk \citep{PR85a,PR85b}.  At such accretion rates, CVs are either in the nova-like state or in the outburst phase of a dwarf nova \citep{warner95,hellier01}.

Alternatively, we can directly compare optical and soft X-ray fluxes, as in the {\it ROSAT} survey of CVs by \citet{BT93}. Converting the {\it XMM-Newton}/EPIC flux for SMSS J1305$-$2931 into a {\it ROSAT}/PSPC 0.1--2.4 keV count rate (assuming suitable thermal-plasma models), we estimate a PSPC rate of $\approx$10$^{-3}$ counts s$^{-1}$. Thus, our source falls into the region of the X-ray/optical diagram populated mostly by nova-like CVs and dwarf novae in the bright state \citep[][their Fig.~2]{BT93}. One problem with this kind of diagram is that the soft X-ray flux is highly affected by an uncertain amount of absorption; for example, we could be under-estimating the soft X-ray luminosity of SMSS J1305$-$2931 dramatically if the system is seen almost edge-on, in comparison with the other sources in the {\it ROSAT} sample at lower inclinations. To avoid this problem, following \citet{Mukai17}, we can use a correlation between the 2--10 keV flux (less affected by absorption) and the visual magnitude. The observed 2--10 keV flux of SMSS J1305$-$2931 is $f_{2-10} \approx 1.1 \times 10^{-13}$ erg cm$^{-2}$ s$^{-1}$ while the absorption-corrected value is $f_{2-10} \approx 1.6 \times 10^{-13}$ erg cm$^{-2}$ s$^{-1}$, almost (to within 10\%) independent of the choice of model. Using either of these values in the \citet{Mukai17} diagram confirms that our source is consistent with nova-like CVs or high-state dwarf novae, but is more than an order of magnitude fainter in X-rays than typical polar and intermediate-polar CVs.

The SMSS J1305$-$2931 system is unlikely to be a dwarf nova in outburst, however, as all the 
available optical 
photometry, including that from surveys such as APASS, Pan-STARRS1 and SkyMapper DR1.1, 
indicate that the optical brightness of the system has remained approximately constant on a timescale 
of years without any 
indication of significant outbursts or transitions.  In particular, 
the 
UVOT $v$ data are constant at the $\sim$0.2--0.3 mag level over the $\sim$60~day monitoring period, 
and are consistent with the Faulkes-S photometry taken $\sim$250~d later (Fig.\ \ref{FS_phot_fig2}).  
We also note that \citet{Ofek2012} classify SMSS J1305--2931 as a non-variable source based 
on Palomar Transient Factory photometry.  As regards longer term variations,
while quantitative information is more difficult to obtain from the photographic DSS blue (1978 February 
17), and POSS-I (1958 April 14) images, estimates of the magnitude of the central star relative to nearby 
stars with APASS $B$ magnitudes, again indicates no substantial change in the brightness of the 
central star at the $\sim$0.2--0.3 mag level.  


From the EW(H$\beta$) versus absolute magnitude relation \citep[e.g.][Fig.\ 3]{patterson11}, with the mean value $\langle$EW(H$\beta$)$\rangle$ $\approx$ 4.4 \AA\/ listed in \ref{Spectra},
we infer that $M_V \la 5.5$ mag for SMSS J1305--2931 (the relation saturates for smaller EWs, corresponding to brighter disks). The observed $V \approx 16.5$ mag, corrected by a line-of-sight Galactic reddening E$(B-V) \approx 0.09$ mag \citep{SF11}, then implies a distance $d \ga 1.5$ kpc, surprisingly large for a CV \citep{patterson11}. A similar absolute brightness, and therefore a similar distance, is also consistent with the relation between accretion rate and absolute brightness in Warner (1987). For convenience, we adopt a distance of 2 kpc and where required, scale distance-related quantities by $d_{2{\rm kpc}}$, the distance in units of 2~kpc.  For distances/absolute magnitudes of the order of those discussed here, the orbital period of the system is likely to be $\sim$4--6 hours \citep{warner87,patterson11}. We can also assume a safe upper limit of $d \la 5$ kpc, based on the maximum X-ray and optical luminosities ever observed for any type of CV \citep{warner87,warner95,patterson11,pretorius12}.

We obtain further constraints on the distance from the He {\sc ii} $\lambda$4686\AA\/ emission line. 
The EW(H$\beta$) versus EW(He {\sc ii} $\lambda$4686) diagram of \citet{vPV84} tells us again that
the average strength of the He {\sc ii} $\lambda$4686 measured for SMSS~J1305--2931
is consistent with those of LMXBs (which we have already ruled out), nova-like CVs, and some IPs, but is not consistent with polar CVs. More interestingly, there is a tight correlation between the emitted luminosity in the He {\sc ii} $\lambda 4686$ line and the accretion rate \citep[][their Fig.\ 5]{PR85b}. 
We have already shown that the X-ray luminosity and the nova-like CV interpretation suggest an accretion rate $\dot{M} \sim 10^{18}$ g s$^{-1}$. For that rate, we expect a He {\footnotesize II} $\lambda 4686$ luminosity $\sim$10$^{30}$ erg s$^{-1}$. From our observed spectra, corrected for line-of-sight reddening, we measure a line luminosity $L_{4686} \approx 5 \times 10^{30} d^2_{2{\rm kpc}}$ erg s$^{-1}$, which demonstrates consistency with our previous distance estimates based on the continuum luminosity.

Finally, we note that the soft X-ray luminosity of SMSS 1305--2931 in the 0.5 to 2.0 keV range is between 
$L_{(0.3-2)} \approx 4 \times 10^{31} \times d^2_{2{\rm kpc}}$ erg s$^{-1}$ and $L_{(0.3-2)} \approx a
few  \times 10^{32} \times d^2_{2{\rm kpc}}$ erg s$^{-1}$ (Table \ref{Tab_xray}), depending on the which of the X-ray spectral models discussed above is adopted. These values place the system among the 
most 
luminous non-magnetic CVs in the luminosity function of \citet{pretorius12}, again consistent with our
interpretation of the system as a nova-like CV in an accretion-dominated phase.  
We note that similar spectral models, 
with softer and harder emission components absorbed by different column densities, have also been 
successfully applied to, for example, the {\it XMM-Newton}/EPIC spectrum of UX UMa \citep{pratt04}, 
and to the {\it Chandra}/HETG spectrum of TT Ari \citep{zemko14}, both of which are nova-like CVs.  

In summary, we argue that a consistent interpretation of the SMSS 1305$-$2931 system is as a 
nova-like CV at $d \sim 2$ kpc, accreting mass at a rate $\dot{M} \sim 10^{-8} M_{\odot}$ yr$^{-1}$, 
which produces a bolometric luminosity $L_{\rm {bol}} \sim 10^{35}$ erg s$^{-1}$ 
(assuming a characteristic white 
dwarf radiative efficiency $\eta \approx 10^{-4}$).  More specifically, our modelling of the X-ray
spectra indicates the presence of two thermal-plasma components.  Based on an analogy with other 
highly accreting CVs, the hotter component ($kT \sim 10$ keV), which is absorbed by a high column 
density ($n_{\rm H} \sim 5 \times 10^{22}$ cm$^{-2}$), can be attributed to an optically thin, radiatively 
inefficient, hot boundary layer.   The cooler thermal-plasma component ($kT < 1$ keV), however, most 
likely originates from a more extended region further out (as suggested by its much lower $n_{\rm H}$), 
possibly the result of harder X-ray photons down-scattered in a disk wind, in agreement with the 
interpretation of other nova-like CV systems 
\citep[see][and references therein]{pratt04,balman14,zemko14}.

\section{Are the blobs outflow relics?}

The two blobs in the SMSS J1305--2931 system are accurately symmetrical about the central object and
we have argued that they are associated with the central star.  Nevertheless, we must first consider the 
possibility that the alignment of the blobs might simply have arisen by chance, in which case the blobs 
would most likely be distant galaxies.  We have investigated this possibility by making use of the 
full 7$^\prime$.7 $\times$ 7$^\prime$.7 AAT 
$H$-band image\footnote{This image was chosen because its limiting magnitude is fainter than 
both the DSS-R image and the Pan-STARRS1 $y/i/g$ image stack.} 
(a sub-section of which is shown in Fig.\ \ref{H_phot_fig}), seeking to identify other 
faint low-surface brightness objects of comparable size to the blobs.  
There are at least half a dozen or so objects with characteristics similar to the blobs visible on the frame,
and
in each case the symmetrical location with respect to the central star was inspected.  However, we found that in all 
cases the symmetrical location was blank sky, with no similar object visible in the vicinity.  While not 
definitive, this outcome suggests that the blobs are unlikely to be the result of a chance alignment.

The blobs lie, in projection on the plane of the sky, at a distance of 2.2 $\times$ $d_{2{\rm kpc}}$ pc 
from the central star, and, with
an on-sky diameter of $\sim$10$\arcsec$, they are of order 0.07 $\times$ 
$d_{2{\rm kpc}}$ pc in size.   As demonstrated in previous sections, they emit faint optical/IR continuum 
radiation, with narrow-band photometry suggesting no strong emission-line features.  The fluxes associated with the $g$, $r$, $J$, $H$, 
and $K$ magnitudes of the blobs do not suggest a power-law spectral energy distribution, rather they 
are more consistent with a spectral energy distribution that peaks at wavelengths between $r$ and 
$J$ in log~$F_{\lambda}$ and in log~$\nu F_{\nu}$.  If interpreted as thermal emission, the characteristic temperature would be $\sim$4500 K for both 
components.  The average luminosity for the blobs in the $r$-band is approximately  $5.6 \times 10^{30} \times d^2_{2{\rm kpc}}$ erg s$^{-1}$.

Our interpretation of these features is that they are the {\it relics} of a  
symmetrical bi-polar outflow, 
or accretion-powered jet, that was once present in the SMSS J1305--2931 system.  
Collimated bipolar outflows or jets (highly-collimated fast flows) are known in a variety of stellar 
systems, ranging from young star-forming systems to LMXBs and HMXBs, 
and symbiotic-star binaries (red giant/white dwarf binary systems with active mass transfer).  
In general, the physical processes 
connecting the collimated outflows and accretion disk properties seem to be similar for most 
of these different types of systems.   
Jets are not commonly observed in CV systems.  Nevertheless, \citet{KR08} have used the radio emission properties of the dwarf nova SS~Cyg in outburst to suggest that the data are best explained as synchrotron emission originating from a transient jet \citep[see also][]{miller-jones11, russell+16}.  Moreover, radio observations at 8--12 GHz  of dwarf novae \citep{coppejans16b} and 
nova-like CVs \citep{kording11,coppejans15} have shown that such optically thin synchrotron emission is a common feature for both steady and transient CVs: the preferred interpretation is that the emission 
originates from the formation of transient or steady jets.    

The presence of a collimated outflow, either currently or recently active, in the nova-like CV SMSS J1305$-$2931, perhaps favoured by the existence of a hot, radiatively inefficient boundary layer, is therefore not implausible. At the suggested distance of $\sim$2 kpc, the radio-bright CVs in the sample of \citet{coppejans16b} would have a 10-GHz flux density of only $\sim$0.02--0.2 $\mu$Jy, well below our non-detection limits (see section \ref{radio}). This range of radio fluxes corresponds to radio luminosities $\approx$10$^{24}$--10$^{26}$ erg s$^{-1}$, that is, $\la$10$^{-9}$ times the total power available from mass accretion. As an interesting analogy, the maximum radio luminosity achievable by neutron stars in a bright state (accretion power $\sim$10$^{38}$ erg s$^{-1}$) seems to be $\sim$10$^{29}$ erg s$^{-1}$ \citep{Tudor2017}, which is a similar fraction of the accretion power.

We speculate that {\it the blobs are the optically-thin passively cooling remnants of previously active collimated outflows in the SMSS J1305--2931 system}.  An alternative scenario, that the blobs represent 
the vestigial remains of 
termination shocks formed when outflowing material impacted into the ISM surrounding SMSS J1305--2931, 
seems ruled out given the likely low density of the ISM at the height above the plane of the system 
(1.1 $\times$ $d_{2{\rm kpc}}$).
An important quantity for any viable interpretation of the blobs is 
the timescale for their evolution.  If the timescale is short, then the probability of observing them at the current epoch becomes small, suggesting that another explanation is warranted.  Alternatively, if the timescale is too long, then the question would arise as to why such features aren't more commonly observed.  Attempting a detailed physical model of the blobs 
is beyond the scope the current paper.  We can, however, make a first-order estimate of an evolutionary timescale for the blobs by calculating the sound speed crossing time, $\tau_{s}$, which is the ratio of the size to the sound speed in the blob material.  The sound speed, $c_{s}$, can be estimated as 
$(k T / \mu m)^{1/2}$, which is 4.7 km s$^{-1}$ for T = 4500 K and solar composition.  For a scale-size of 0.07~pc, the value of 
$\tau_{s}$ for the blobs is then $\sim$15,000 yrs $\times$ $d_{2{\rm kpc}}$.  This is long enough that we are not required to be observing the system at a particularly special epoch, but short enough that similar systems should be quite rare, as seems to be the case.

The principle alternative explanation of the blobs is that they are the last remnants of a
nova eruption that occurred in the system in the historical past.  A potentially related example is the
nearby (d $\approx$ 160 pc) dwarf nova Z~Cam.  \citet{Shara2007} have detected a number of 
thin low surface-brightness arc-like structures surrounding the central object in this system.  
The features, which lie
$\sim$0.7 pc from the central source, are interpreted as the result of a classical nova 
eruption of the system that occurred a few thousand years ago.  \citet{Shara2007} argue that the
nova ejecta swept up surrounding interstellar material to form the shock-ionised arcs seen at the present-day.  The arcs in the Z~Cam system differ from the blobs in the SMSSJ1305--2931 system in that their spectra show definite emission lines of H$\alpha$ and [N{\sc ii}] and [O{\sc ii}] \citep{Shara2007}.  Yet it is conceivable that a classical nova (or a recurrent nova -- see \citet{Darnlet17}) eruption also occurred in the SMSSJ1305--2931 system in the past, and that the blobs are the last remaining features from that eruption.  \citet{Mason18} point out in their detailed study of four novae that the ejecta properties
are in all cases consistent with ``clumpy gas expelled during a single, brief ejection episode and in ballistic expansion'' confined within a bi-conical geometry \citep{Mason18}.
Assuming an outflow velocity of order 1000 km s$^{-1}$, or more \citep[e.g.][]{Yaron2005}, the projected radial distance for the SMSSJ1305--2931 blobs of 2.2 $\times$ $d_{2{\rm kpc}}$ pc implies a timescale of 2000 $d_{2{\rm kpc}}$ yrs, or less, since the 
eruption.  In that context it is also worth noting that at least some novae have been shown to have 
complex post-eruption structures, including
multiple components, collimated outflows and high-velocity features \citep[e.g.][]{Darnlet17,Woudt09}.
Further, \citet{Mroz16} have shown that after a nova eruption, the white dwarf has a higher accretion rate
than before the nova outburst, for a period of order hundreds of years.  We have argued that
SMSSJ1305--2931 is a nova-line CV, i.e.\ has a high accretion rate, and that is then consistent with an
interpretation of the system as being in a post-nova state.

\section{Summary}

We report here the serendipitous discovery of a probable nova-like CV, designated 
SMSS J130522.47--293113.0.  The system is a weak X-ray source, and is not detected at radio wavelengths.   The optical spectrum of the central object is characteristic of radiation from a hot accretion 
disk surrounding a compact companion in a close-binary system.  It is dominated by emission lines of hydrogen, neutral and ionized helium, and the N {\sc iii}, C {\sc iii} complex.  The lines vary significantly in strength, width, and velocity on timescales at least short as 1 day: the velocity variations have scales of 100's km~s$^{-1}$ and the line widths vary from $\sim$300 to $\sim$1000 km~s$^{-1}$.  Broad-band optical photometry also reveals significant low-level variations in the output of the system at the 0.2--0.3 mag level, with timescales as short as 10-15 min, as well as over days and weeks.  More extensive long-term monitoring of the system is required in order to constrain the period of the binary orbit which we estimate to be approximately 5 hours, based on the likely distance ($\sim$2 kpc) and absolute magnitude (M$_{V}$ $\approx$ 4.7) of the system.

The most intriguing feature of the environment of the SMSS J1305--2931 system is the presence of two small (size $\sim$0.07pc) low surface brightness blobs that are symmetrically located $\sim$2 pc (on the plane of the sky) either side of the central object.  The blobs are detected only at optical and near-IR wavelengths, and their optical/IR emission appears to be dominated by continuum radiation.  Our preferred
interpretation is that the blobs are the relics of previous accretion-powered collimated outflows in the system: as such 
they are likely the first known example of remnants of accretion-powered collimated outflows in a stellar system.  While it would undoubtedly be a challenging observation even with the largest telescopes (blob surface brightness $\sim$ 24.7 $g$-mag per arcsec$^2$ on a $\sim$6 arcsec
spatial scale), low-resolution spectroscopy of the blobs would be useful to confirm the apparent lack of any significant emission-line flux, and to assist in the clarification their nature.

\section*{Acknowledgements}

We thank Axel Schwope, Geoff Bicknell and the referee for useful comments and suggestions.  RS acknowledges support from a Curtin University Senior Research Fellowship. GDC would like to acknowledge research support from the Australian Research Council (ARC) through Discovery Grant programs DP120101237 and DP150103294.  BPS acknowledges support from the ARC through the Laureate Fellowship LF0992131.  This research was also supported in part by the Australian Research Council Centre of Excellence for All-sky Astrophysics (CAASTRO) through project number CE110001020.
TCB acknowledges partial support from grant PHY 14-30152, Physics Frontier Center/JINA Center for the Evolution of the Elements (JINA-CEE), awarded by the US National Science Foundation. 
  
The national facility capability for SkyMapper has been funded through ARC LIEF grant LE130100104 from the Australian Research Council, awarded to the University of Sydney, the Australian National University, Swinburne University of Technology, the University of Queensland, the University of Western Australia, the University of Melbourne, Curtin University of Technology, Monash University and the Australian Astronomical Observatory. SkyMapper is owned and operated by The Australian National University's Research School of Astronomy and Astrophysics. The survey data were processed and provided by the SkyMapper Team at ANU\@. The SkyMapper node of the All-Sky Virtual Observatory
(ASVO) is hosted at the National Computational Infrastructure (NCI).  Development and support the SkyMapper node of the ASVO has been funded in part by Astronomy Australia Limited (AAL) and the Australian Government through the Commonwealth's Education Investment Fund (EIF) and National Collaborative Research Infrastructure Strategy (NCRIS), particularly the National eResearch Collaboration Tools and Resources (NeCTAR) and the Australian National Data Service Projects (ANDS).

The Australia Telescope is funded by the Commonwealth of Australia for operation as a National Facility managed by CSIRO.

This research includes observations obtained at the Southern Astrophysical Research (SOAR) telescope, 
which is a joint  project of the Minist\'{e}rio da Ci\^{e}ncia, Tecnologia, e Inova\c{c}\~{a}o (MCTI) 
da Rep\'{u}blica Federativa do Brasil, the U.S. National Optical Astronomy Observatory (NOAO), the 
University of North Carolina at Chapel Hill (UNC), and Michigan State University (MSU).

The Digitized Sky Surveys were produced at the Space Telescope Science Institute under U.S. 
Government grant NAG W-2166. The images of these surveys are based on photographic data obtained 
using the Oschin Schmidt Telescope on Palomar Mountain and the UK Schmidt Telescope. The plates 
were processed into the present compressed digital form with the permission of these institutions.
The National Geographic Society - Palomar Observatory Sky Atlas (POSS-I) was made by the California 
Institute of Technology with grants from the National Geographic Society.
The Second Palomar Observatory Sky Survey (POSS-II) was made by the California Institute of 
Technology with funds from the National Science Foundation, the National Geographic Society, the Sloan 
Foundation, the Samuel Oschin Foundation, and the Eastman Kodak Corporation.  The Oschin Schmidt 
Telescope is operated by the California Institute of Technology and Palomar Observatory.  The UK 
Schmidt Telescope was operated by the Royal Observatory Edinburgh, with funding from the UK Science 
and Engineering Research Council (later the UK Particle Physics and Astronomy Research Council), until 
1988 June, and thereafter by the Anglo-Australian Observatory.  The blue plates of the southern Sky Atlas 
and its Equatorial Extension (together known as the SERC-J), as well as the Equatorial Red (ER), and the 
Second Epoch [red] Survey (SES) were all taken with the UK Schmidt.  Supplemental funding for sky-
survey work at the ST ScI is provided by the European Southern Observatory.

This research has also made use of the VizieR catalogue access tool, CDS, Strasbourg, France.   
The original description of the Vizier service was published in A\&AS, 143, 23.

\end{document}